\documentclass{article}

\usepackage{amssymb}
\newcommand{\be}{\begin{equation}}
\newcommand{\ee}{\end{equation}}
\def\bea{\begin{eqnarray}}
\def\eea{\end{eqnarray}}
\def\der{\partial}
\newcommand{\ov}{\overline}

\newcommand{\la}{\langle}
\newcommand{\ra}{\rangle}

\newcommand{\wh}{\widehat}
\newcommand{\mscr}[1]{\mbox{\scriptsize #1}}

\newcommand{\ft}[2]{{\textstyle\frac{#1}{#2}}}

\def\ve{\varepsilon}

\renewcommand{\d}{\delta}

\newcommand{\g}{\gamma}

\newcommand{\m}{\mu}

\newcommand{\n}{\nu}

\newcounter{exercise}
\refstepcounter{exercise}

\newtheorem{exercises}[exercise]{{\bf Exercise}}
\newenvironment{Exercise}{\begin{exercises}: \sf}{\end{exercises}}
\newcounter{solution}
\refstepcounter{solution}

\newtheorem{solutions}[solution]{{\bf Solution}}
\newenvironment{Solution}{\begin{solutions}: \rm}{\end{solutions}}
\begin{document}

\begin{titlepage}
\begin{center}
\hfill {\tt hep-th/0004098}\\

\vskip 3cm

{ \LARGE \bf Black Holes in Supergravity and String Theory}

\vskip .3in

{\bf Thomas Mohaupt\footnote{\mbox{
\tt 
mohaupt@hera1.physik.uni-halle.de}}}
\\

\vskip 1cm

{\em 
\centerline{Martin-Luther-Universit\"at Halle-Wittenberg, 
Fachbereich Physik,
D-06099 Halle, Germany}    }

\vskip .1in
\end{center}

\vskip .2in

\begin{center} {\bf ABSTRACT } \end{center}
We give an elementary introduction to black holes in 
supergravity and string theory.\footnote{Based on lectures given at
the school of the TMR network 'Quantum aspects of gauge
theories, supersymmetry and unification' in Torino,
January 26 - February 2, 2000.} The focus is on the
role of BPS solutions in four- and higher-dimensional 
supergravity and in string theory. Basic ideas and techniques
are explained in detail, including exercises with solutions.


\vfill

March 2000\\
\end{titlepage}


\tableofcontents

\section{Introduction}

String theory has been the leading candidate for a unified
quantum theory of all interactions during the last 15 years.
The developements of the last five years have 
opened the possibility to go beyond perturbation theory and to
address the most interesting problems of quantum gravity. Among
the most prominent of such problems are those related to black holes:
the interpretation of the Bekenstein-Hawking entropy, Hawking
radiation and the information problem. 

The present set of lecture notes aims to give a paedagogical
introduction to the subject of black holes in
supergravity and string theory. It is primarily intended for
graduate students who are interested in black hole physics,
quantum gravity or string theory. No particular  previous knowledge of these
subjects is assumed, the notes should be accessible for any 
reader with some background in general relativity and quantum 
field theory. The basic ideas and techniques are treated 
in detail, including exercises and their solutions.
This includes 
the definitions of mass, surface gravity and entropy of black holes, the
laws of black hole mechanics, the interpretation of the extreme
Reissner-Nordstrom black hole as a supersymmetric soliton,
$p$-brane solutions of higher-dimensional supergravity, their
interpretation in string theory and their relation to $D$-branes,
dimensional reduction of supergravity actions, 
and, finally, the construction of extreme black holes by dimensional
reduction of $p$-brane configurations. Other topics, which are needed
to make the lectures self-contained are explained in a summaric 
way. Busher $T$-duality is mentioned briefly and studied further 
in some of the exercises. Many other topics are omitted, 
according to the motto 'less is more'.

A short commented list of references is given at the end of 
every section. It is not intended to provide a representative or 
even full account of the literature, but to give suggestions for
further reading. Therefore we recommend, based on subjective 
preference, some books, reviews and research papers.

\section{Black holes in Einstein gravity}
\setcounter{equation}{0}

\subsection{Einstein gravity}

The basic idea of Einstein gravity is that the geometry of
space-time is dynamical and is determined by the distribution of
matter. Conversely the motion of matter is determined by the
space-time geometry: In absence of non-gravitational forces
matter moves along geodesics.

More precisely space-time is taken to be a (pseudo-) Riemannian
manifold with metric $g_{\m\n}$. Our choice of signature is
$(-+++)$. The reparametrization-invariant properties of the metric
are encoded in the Riemann curvature tensor $R_{\m \n \rho \sigma}$, 
which is related by the
gravitational field equations to the energy-momentum tensor of
matter, $T_{\m \n}$. If one restricts the action to be at most
quadratic in derivatives, and if one ignores the possibility of
a cosmological constant,\footnote{We will set the cosmological constant
to zero throughout.} then the unique gravitational action 
is the Einstein-Hilbert action,
\be
S_{EH} = \frac{1}{2 \kappa^2} \int \sqrt{-g} R\;,
\ee 
where $\kappa$ is the gravitational constant, which will be related
to Newton's constant below. The coupling to matter is determined
by the principle of minimal coupling, i.e. one replaces partial
derivatives by covariant derivatives with respect to the Christoffel
connection $\Gamma_{\m\n}^{\rho}$.\footnote{In the case of fermionic
matter one uses the vielbein $e_{\m}^{\;\;a}$ instead of the
metric and one introduces a second connection, the spin-connection
$\omega_{\m}^{ab}$, to which the fermions couple.}

The energy-momentum tensor of matter is 
\be
T_{\m\n} = \frac{-2}{\sqrt{-g}} \frac{\d S_M}{\d g^{\m \n}} \,,
\ee
where $S_M$ is the matter action. The Euler-Lagrange equations
obtained from variation of the combined action $S_{EH}+S_M$ with
respect to the metric are the Einstein equations
\be
R_{\m\n} - \ft12 g_{\m\n} R = \kappa^2 T_{\m \n} \;.
\label{EinsteinEq}
\ee
Here $R_{\m \n}$ and $R$ are the Ricci tensor and the Ricci scalar,
respectively.


The motion of a massive point particle in a given space-time background
is determined by the equation
\be
m a^\n = m \dot{x}^\m \nabla_\m \dot{x}^\n = 
m \left( \ddot{x}^\n + \Gamma^\n_{\m \rho} \dot{x}^\m \dot{x}^{\rho} \right)
= f^\n \;,
\label{forcelawgr}
\ee
where $a^\n$ is the acceleration four-vector, 
$f^\n$ is the force  four-vector of non-gravitational forces and
$\dot{x}^{\m} = \ft{dx^\m}{d \tau}$ is the derivative with respect to
proper time $\tau$.
 
In a flat background or in a local
inertial frame equation (\ref{forcelawgr}) 
reduces to the force law of special relativity,
$m \ddot{x}^\n = f^\n$. If no (non-gravitational) forces are present,
equation (\ref{forcelawgr}) becomes the geodesic equation,
\be
\dot{x}^\m \nabla_\m \dot{x}^\n = \ddot{x}^\n + \Gamma^\n_{\m \rho}
\dot{x}^\m \dot{x}^{\rho} = 0 \;. 
\label{GeodesicEq}
\ee

One can make contact with Newton gravity by considering the
Newtonian limit. This is the limit of small curvature and 
non-relativistic velocities $v \ll 1$ (we take $c=\hbar=1$). 
Then the metric can be expanded around the Minkowski metric
\be
g_{\m \n} = \eta_{\m \n} + 2 \psi_{\m\n} \;,
\ee
where $|\psi_{\m \n}| \ll 1$. If this expansion is carefully performed in
the Einstein equation (\ref{EinsteinEq})
and in the geodesic equation (\ref{GeodesicEq}) one finds
\be
\Delta V = 4 \pi G_N \rho \;\;\; \mbox{and} \;\;\;
\frac{d^2 \vec{x}}{dt^2} = - \vec{\nabla} V \;,
\ee
where $V$ is the Newtonian potential, $\rho$ is the matter density,
which is the leading part of $T_{00}$, and $G_N$ is Newton's
constant. The proper time $\tau$ has been eliminated in terms
of the coordinate time $t=x^0$.
Thus one gets the potential equation for Newton's gravitational 
potential and the equation of motion for a point
particle in it.
The Newtonian potential $V$ and Newton's constant $G_N$ are related
to $\psi_{00}$ and $\kappa$ by
\be
V = - \psi_{00} \mbox{   and   } \kappa^2 = 8 \pi G_N \;.
\label{matching}
\ee

In Newtonian gravity a point mass or spherical mass distribution
of total mass $M$ gives rise to a potential $V=-G_N \ft{M}{r}$.
According to (\ref{matching}) this corresponds to a leading
order deformation of the flat metric of the form
\be
g_{00} = - 1 + 2 \frac{G_N M}{r} + O(r^{-2}) \;.
\label{DefMass}
\ee
We will use equation (\ref{DefMass}) as our working definition for the 
mass of an asymptotically flat space-time. 
Note
that there is no natural way to define the mass of a general space-time
or of a space-time region. 
Although we have a local conservation law
for the energy-momentum of matter, $\nabla^\m T_{\m \n}=0$, 
there is in general no way to construct a reparametrization invariant
four-momentum
by integration because $T_{\m\n}$ is a symmetric tensor. 
Difficulties in defining a meaningful conserved 
mass and four-momentum for a general space-time are 
also expected for a second reason. The principle 
of equivalence implies that the gravitational field can be
eliminated locally by going to an inertial frame. Hence,
there is no local energy density associated with gravity.
But since the concept of mass works well in Newton gravity and
in special relativity, we expect that one can define the mass
of isolated systems, in particular the mass of an asymptotically
flat space-time. Precise definitions can be given by different 
constructions, like the ADM mass and the Komar mass. More generally
one can define the four-momentum and the angular momentum of an
asymptotically flat space-time.

For practical purposes it is convenient to extract the mass by
looking for the leading deviation of the metric from flat space,
using (\ref{DefMass}).
The quantity $r_S = 2 G_N M$ appearing in the metric
(\ref{DefMass}) has the dimension of a length and is called
the Schwarzschild radius. From now on we will use Planckian units
and set $G_N=1$ on top of $\hbar=c=1$, unless dimensional
analysis is required.

\subsection{The Schwarzschild black hole}

Historically,
the Schwarzschild solution was the first exact solution to Einstein's
ever found. According to Birkhoff's theorem it is the unique spherically
symmetric vacuum solution.

Vacuum solutions are those with a vanishing engergy momentum tensor,
$T_{\mu \nu}=0$. By taking the trace of Einsteins equations this implies
$R=0$ and as a consequence 
\be
R_{\m \n} = 0 \;.
\ee
Thus the vacuum solutions to Einsteins equations are precisely the
Ricci-flat space-times. 

A metric is called spherically symmetric if it has a group of spacelike
isometries with compact orbits which is isomorphic to the rotation
group $SO(3)$. One can then go to adapted coordinates 
$(t,r,\theta,\phi)$, where $t$ is time, $r$ a radial variable and
$\theta,\phi$ are angular variables, such that the metric takes the
form 
\be
ds^2 = - e^{2f(t,r)} dt^2 + e^{2g(t,r)} dr^2 + r^2 d\Omega^2 \;,
\label{sphericmetric}
\ee
where $f(t,r),g(t,r)$ are arbitrary functions of $t$ and $r$ and
$d \Omega^2 = d \theta^2 + \sin^2 \theta d\phi^2$ is the 
line element on the unit two-sphere.

According to Birkhoff's theorem the Einstein
equations determine the functions $f,g$ uniquely. In particular
such a solution must be static. A metric is called stationary
if it has a timelke isometry. If one uses the integral lines
of the corresponding Killing vector field to define the time coordinate
$t$, then the metric is $t$-independent, $\der_t g_{\m \n} =0$.
A stationary metric is called static if in addition the timelike Killing
vector field is hypersurface orthogonal, which means that it is
the normal vector field of a family of hypersurfaces. In this case
one can eliminate the mixed components $g_{ti}$ of the metric by
a change of coordinates.\footnote{In (\ref{sphericmetric}) these
components have been eliminated using spherical symmetry.}

In the case of a general spherically symmetric metric (\ref{sphericmetric})
the Einstein equations determine the functions $f,g$ 
to be $e^{2f} = e^{-2g} = 1 - \ft{2M}{r}$. This is the Schwarzschild
solution:
\be
ds^2 = - \left(1 - \frac{2M}{r} \right) dt^2 
+ \left(1 - \frac{2M}{r} \right)^{-1} dr^2 + r^2 d\Omega^2 \;.
\label{Schwarzschild}
\ee
Note that the solution is asymptitotically flat,
$g_{\m\n}(r) \rightarrow_{r \rightarrow \infty} \eta_{\m\n}$.
According to the discussion of the last section, $M$ is the
mass of the Schwarzschild space-time.

One obvious feature of the Schwarzschild metric is that it becomes
singular at the Schwarzschild radius $r_S = 2M$, where
$g_{tt}=0$ and $g_{rr} = \infty$. Before investigating this further
let us note that $r_S$ is very small: For the sun one finds
$r_S = 2.9 \mbox{km}$ and for the earth $r_S = 8.8 \mbox{mm}$.
Thus for atomic matter the Schwarzschild radius is inside the matter 
distribution. Since the Schwarzschild solution is a vacuum solution,
it is only valid outside the matter distribution. Inside one has
to find another solution with the energy-momentum tensor
$T_{\m\n} \not=0$ describing the system under consideration and
one has to glue the two solutions at the boundary. The singularity of
the Schwarzschild metric at $r_S$ has no significance in this case.
The same applies to nuclear matter, i.e. neutron
stars. But stars with a mass above the Oppenheimer-Volkov
limit of about 3 solar masses are instable against total gravitational
collapse. If such a collapse happens in a spherically symmetric way,  
then the final state must be the Schwarzschild metric, as a 
consequence of Birkhoff's theorem.\footnote{The assumption of a spherically
symmetric collapse might seem unnatural. We will not discuss 
rotating black holes in these lecture notes, but there is a generalization
of Birkhoff's theorem with the result that the most general uncharged
stationary black hole solution in Einstein gravity is the Kerr black
hole. A Kerr black hole is uniquely characterized by its mass and 
angular momentum.  
The stationary final state of an arbitrary collapse 
of neutral matter in Einstein gravity 
must be a Kerr black hole. Moreover rotating black holes,
when interacting with their environment, rapidly loose angular
momentum by superradiance. In the limit of vanishing angular momentum 
a Kerr black hole becomes a Schwarzschild black hole.
Therefore even a non-spherical collapse 
of neutral matter can have a Schwarzschild black hole
as its (classical) final state.} 
In this situation the question of
the singularity of the Schwarzschild metric at $r=r_S$ becomes
physically relevant. As we will review next, $r=r_S$ is a so-called
event horizon, and the solution describes a black hole. There is
convincing observational evidence that such objects exist.

We now turn to the question what happens at $r=r_S$. One observation
is that the singularity of the metric is a coordinate singularity,
which can be removed by going to new coordinates, for example
to Eddington-Finkelstein or to Kruskal coordinates. As a consequence
there is no curvature singularity, i.e. any coordinate invariant
quantity formed out of the Riemann curvature tensor is finite. In
particular the tidal forces on any observer at $r=r_S$ are finite
and even arbitrarily small if one makes $r_S$ sufficiently large.
Nevertheless the surface $r=r_S$ is physically distinguished: It is
a future event horizon. This property can be characterized in
various ways. 

Consider first the free radial motion of a massive particle 
(or of a local observer in a starship) between positions
$r_2 > r_1$. Then the time $\Delta t = t_1 - t_2$ needed to travel
from $r_2$ to $r_1$ diverges in the limit $r_1 \rightarrow r_S$:
\be
\Delta t \simeq r_S \log \frac{ r_2 - r_S }{ r_1 - r_S } 
\rightarrow_{r_1 \rightarrow r_S} \infty \;.
\ee
Does this mean that one cannot reach the horizon? Here we have
to remember that the time $t$ is the coordinate time, i.e.
a timelike coordinate that we use to label events. It is 
not identical with the time measured by a freely falling observer.
Since the metric is asymptotically flat, the Schwarzschild 
coordinate time coincides with the proper time of an observer
at rest at infinity. Loosely speaking an observer at infinity
(read: far away from the black hole) never 'sees' anything reach
the horizon. This is different from the perspective of a
freely falling observer. For him the difference $\Delta \tau = \tau_1
- \tau_2$ of proper time is finite:
\be
\Delta \tau = \tau_1 - \tau_2 = \frac{2}{3 \sqrt{3 r_S}}
\left( r_2^{3/2} - r_1^{3/2} \right)
\rightarrow_{r_1 \rightarrow r_S}  \frac{2}{3 \sqrt{3 r_S}}
\left( r_2^{3/2} - r_S^{3/2} \right) \;.
\ee
As discussed above the gravitational forces at $r_S$ are finite
and the freely falling observer will enter the inerior region
$r<r_S$. The consequences will be considered below.

Obviously the proper time of the freely falling observer
differs the more from the Schwarzschild time the closer he gets
to the horizon. The precise relation between the infinitesimal
time intervals is
\be
\frac{d \tau}{dt}  = \sqrt { - g_{tt} } = \left( 1 - \frac{r_S}{r}
\right)^{1/2} =: V(r) \;.
\ee
The quantity $V(r)$ is called the redshift factor associated with the
position $r$. This name is motivated by our second thought experiment.
Consider two static observers at positions $r_1 < r_2$. 
The observer at $r_1$ emits a light ray of frequency $\omega_1$
which is registered at $r_2$ with frequency $\omega_2$. The
frequencies are related by
\be
\frac{\omega_1}{\omega_2} = \frac{V(r_2)}{V(r_1)} \,.
\label{redshift}
\ee
Since $\ft{V(r_2)}{V(r_1)} < 1$, a lightray which travels outwards
is redshifted, $\omega_2 < \omega_1$. Moreover, since
the redshift factor vanishes at the horizon, $V(r_1 = r_S)=0$,
the frequency $\omega_2$ goes to zero, if the
source is moved to the horizon. Thus, the event horizon can
be characterized as a surface of infinite redshift.

\begin{Exercise}
Compute the Schwarzschild time that a lightray needs in order to
travel from $r_1$ to $r_2$. What happens in the limit 
$r_1 \rightarrow r_S$?
\end{Exercise}

\begin{Exercise}
Derive equation (\ref{redshift}). \\ {\bf Hint 1:}
If $k_\m$ is the four-momentum of the lightray and if
$u^\m_i$ is the four-velocity of the static observer at
$r_i$, $i=1,2$, then the frequency measured in the frame
of the static observer is
\be
\omega_i = - k_\m u^\m_i \;.
\ee
(why is this true?). \\ {\bf Hint 2:} If $\xi^\m$ is a Killing vector
field and if $t^\m$ is the tangent vector to a geodesic, then
\be
t^\m \nabla_\m ( \xi_\n k^\n ) =  0 \;,
\label{EnergyConservation}
\ee
i.e. there is a conserved quantity. (Proof this. What is the
meaning of the conserved quantity?) \\{\bf Hint 3:} What is the relation
between $\xi^\m$ and $u^\m_i$?
\end{Exercise}

Finally, let us give a third characterization of the event horizon.
This will also enable us to introduce a quantity called the surface
gravity, which will play an important role later. Consider a
static observer at position $r>r_S$ in the Schwarzschild space-time.
The corresponding world line is not a geodesic and therefore
there is a non-vanishing accelaration $a^\m$. In order to
keep a particle (or starship) of mass $m$ at position, a 
non-gravitational force $f^\m=ma^\m$ must act according to
(\ref{forcelawgr}). For a Schwarzschild space-time the acceleration
is computed to be
\be
a^\m = \nabla^\m \log V(r)
\label{amu}
\ee
and its absolute value is
\be
a = \sqrt{a^\m a_\m}
= \frac{ \sqrt{ \nabla_\m V(r) \nabla^\m V(r)} }{V(r)} \;.
\label{a}
\ee
Whereas the numerator is finite at the horizon
\be
\sqrt{ \nabla_\m V(r) \nabla^\m V(r)} = \frac{r_S}{2r^2}
\rightarrow_{r \rightarrow r_S} \frac{1}{2 r_S} \;,
\ee
the denominator, which is just the redshift factor, goes to 
zero and the acceleration diverges. Thus the event horizon is a 
place where one cannot keep position. 
The finite quantity
\be
\kappa_S := ( V a)_{r=r_S}
\ee
is called the surface gravity of the event horizon. This quantity
characterizes the strength of the gravitational field. For a
Schwarzschild black hole we find
\be
\kappa_S= \frac{1}{2r_S} = \frac{1}{4M} \;.
\label{SurGraSch}
\ee

\begin{Exercise}
Derive (\ref{amu}), (\ref{a}) and (\ref{SurGraSch}).
\end{Exercise}

Summarizing we have found that the interior region $r<r_S$ can
be reached in finite proper time from the exterior but is 
causally decoupled in the sense that no matter or light can
get back from the interior to the exterior region. The future
event horizon acts like a semipermeable membrane which can only
be crossed from outside to inside.\footnote{In the opposite case
one would call it a past event horizon and the corresponding 
space-time a white hole.}

Let us now briefely discuss what happens in the interior region.
The proper way to proceed is to introduce new coordinates, which are
are regular at $r=r_S$ and then to analytically continue to $r<r_S$.
Examples of such coordintes are 
Eddington-Finkelstein or Kruskal coordinates. But it turns out
that the interior region $0<r<r_S$ of the Schwarzschild metric
(\ref{Schwarzschild})
is isometric to the corresponding region of the analytically continued
metric. Thus we might as well look at the Schwarzschild metric
at $0<r<r_S$. And what we see is suggestive: the terms $g_{tt}$ and
$g_{rr}$ in the metric flip sign, which says that 
'time' $t$ and 'space' $r$ exchange their roles.\footnote{Actually
the situation is slightly asymmetric between $t$ and $r$. $r$ is
a good coordinate both in the exterior region $r>r_S$ and interior
region $r<r_S$. On the other hand $t$ is a coordinate in the exterior
region, and takes its full range of values $- \infty < t < \infty$
there. The associated timelike Killing vector field becomes
lightlike on the horizon and spacelike in the interior. One can
introduce a spacelike coordinate using its integral lines, and
if one calls this coordinate $t$, 
then the metric takes the form of a Schwarzschild
metric with $r<r_S$. But note that the 'interior $t$' is not the
the analytic extension of the Schwarzschild time, whereas $r$
has been extended analytically to the interior.}
In the interior
region $r$ is a timelike coordinate and every timelike or
lightlike geodesic has to proceed to smaller and smaller
values of $r$ until it reaches the point $r=0$. One can show
that every timelike geodesic reaches this point in finite proper
time (whereas lightlike geodesics reach it at finite 'affine parameter',
which is the substitute of proper time for light rays).

Finally we have to see what happens at $r=0$. The metric becomes
singular but this time 
the curvature scalar diverges, which shows that there is a
curvature singularity. Extended objects are subject to infinite
tidal forces when reaching $r=0$. It is not possible to
analytically continue geodesics beyond this point.

\subsection{The Reissner-Nordstrom black hole}

We now turn our attention to Einstein-Maxwell theory. The action
is
\be
S = \int d^4 x \sqrt{-g} \left( \frac{1}{2 \kappa^2} R -
\frac{1}{4} F_{\m \n} F^{\m \n} \right) \;.
\ee
The curved-space Maxwell equations are the combined set
of the Euler-Lagrange equations and Bianchi identities 
for the gauge fields:
\bea
\nabla_\m F^{\m \n} &=& 0 \;,\\
\ve^{\m \n \rho \sigma} \der_\n F_{\rho \sigma} &=& 0 \;. \\
\nonumber
\eea
Introducing the dual gauge field
\be
{}^{\star} F_{\m \n} = \frac{1}{2} \ve_{\m \n \rho \sigma}
F^{\rho \sigma} \;,
\label{dualgaugefield}
\ee
one can rewrite the Maxwell equations in a more symmetric way, either
as
\be
\nabla_\m F^{\m \n} = 0 \mbox{   and   }
\nabla_\m {}^{\star} F^{\m \n} = 0 
\ee
or as
\be
\ve^{\m \n \rho \sigma} \der_\n {}^{\star} F_{\rho \sigma} = 0
\mbox{  and  }
\ve^{\m \n \rho \sigma} \der_\n F_{\rho \sigma} = 0 \,.
\ee
In this form it is obvious that the Maxwell equations are 
invariant under duality transformations
\be
\left( \begin{array}{c}
F_{\m \n} \\ {}^{\star} F_{\m\n} \\
\end{array} \right)  \longrightarrow
\left( \begin{array}{cc}
a & b \\ c & d \\
\end{array} \right)
\left( \begin{array}{c}
F_{\m \n} \\ {}^{\star} F_{\m\n} \\
\end{array} \right) \;, \mbox{  where   }
\left( \begin{array}{cc}
a & b \\ c & d \\
\end{array} \right) \in GL(2,\mathbb{R}) \;. 
\ee
These transformations include electric-magnetic duality transformations
$F_{\m \n} \rightarrow {}^{\star}F_{\m \n}$. Note that duality transformations
are invariances of the field equations but not of the action.

In the presence of source terms the Maxwell equations 
are no longer invariant under continuous duality transformations.
If both electric and magnetic charges exist, one can still have
an invariance. But according to the Dirac quantization condition
the spectrum of electric and magnetic charges is discrete and the
duality group is reduced to a discrete subgroup of $GL(2,\mathbb{R})$.

Electric and magnetic charges $q,p$ can be written as surface
integrals,
\be
q = \frac{1}{4 \pi} \oint {}^{\star} F \;, \;\;\,
p = \frac{1}{4 \pi} \oint F \;,
\label{Defcharges}
\ee
where $F = \frac{1}{2} F_{\m \n} dx^\m dx^\n$ is the
field strength two-form and the integration surface surrounds
the sources. Note that the integrals have a reparametrization
invariant meaning because one integrates a two-form. This was different
for the mass.

\begin{Exercise}
Solve the Maxwell equations in a static and spherically symmetric
background,
\be
ds^2 = - e^{2g(r)} dt^2 + e^{2f(r)} dr^2 + r^2 d\Omega^2
\label{MetricStaticSpheric}
\ee
for a static and spherically symmetric gauge field.
\end{Exercise}

We now turn to the gravitational field equations,
\be
R_{\m \n} - \frac{1}{2} g_{\m \n} R = \kappa^2
\left( F_{\m \rho} F^{\rho}_{\;\;\n} - \frac{1}{4} g_{\m \n}
F_{\rho \sigma} F^{\rho \sigma} \right) \,.
\label{EMequation}
\ee
Taking the trace we get $R=0$. This is always the case if the
energy-momentum tensor is traceless.

There is a generalization of Birkhoff's theorem: The unique
spherically symmetric solution of (\ref{EMequation}) is
the Reissner-Nordstrom solution
\be
\begin{array}{l}
ds^2 = - e^{2f(r)} dt^2 + e^{-2f(r)} dr^2 + r^2 d \Omega^2 \\
F_{tr} = - \frac{q}{r^2} \;, \;\;\;
F_{\theta \phi} = p \sin \theta \\
e^{2f(r)} = 1 - \frac{2M}{r} + \frac{q^2 + p^2}{r^2} \\
\end{array}
\ee
where $M,q,p$ are the mass and the electric and magnetic charge.
The solution is static and asymptotically flat.

\begin{Exercise}
Show that $q,p$ are the electric and magnetic charge, as
defined in (\ref{Defcharges}).
\end{Exercise}

\begin{Exercise}
Why do the electro-static field $F_{tr}$ and the
magneto-static field $F_{\theta \phi}$ look so different?
\end{Exercise}

Note that it is sufficient to know the electric Reissner-Nordstrom
solution, $p=0$. The dyonic generalization can be generated by
a duality transformation.

We now have to discuss the Reissner-Nordstrom metric. 
It is convenient to rewrite
\be
e^{2f}= 1 - \frac{2M}{r} + \frac{Q^2}{r^2}
= \left( 1 - \frac{r_+}{r} \right) \left( 1 - \frac{r_-}{r} \right) \;,
\ee
where we set $Q = \sqrt{q^2 + p^2}$ and
\be
r_{\pm} = M \pm \sqrt{M^2 - Q^2} \;.
\ee
There are three cases to be distinguished:
\begin{enumerate}
\item
$M>Q>0$: The solution has two horizons, an event horizon
at $r_+$ and a so-called Cauchy horizon at $r_-$. This is
the non-extreme Reissner-Nordstrom black hole. The surface
gravity is $\kappa_S = \frac{r_+ - r_-}{2 r_+^2}$.
\item
$M=Q>0$: In this limit the two horizons coincide at
$r_+=r_- = M$ and the mass equals the charge. This is the
extreme Reissner-Nordstrom black hole. The surface gravity
vanishes, $\kappa_S=0$.
\item
$M<Q$: There is no horizon and the solution has a naked 
singularity. Such solutions are believed to be unphysical.
According to the cosmic censorship hypothesis the only physical 
singularities
are the big bang, the big crunch, and singularities hidden behind
event horizons, i.e. black holes. 
\end{enumerate}

\subsection{The laws of black hole mechanics}

We will now discuss the laws of black hole mechanics. This is
a remarkable set of relations, which is formally equivalent
to the laws of thermodynamics. The significance of this will
be discussed later. Before we can formulate the laws, we need
a few definitions. 

First we need to give a general definition of a black hole and of 
a (future) event horizon. Intuitively a black hole is a region
of space-time from which one cannot escape. In order make the term
'escape' more precise, one considers the behaviour of time-like
geodesics. In Minkowski space all such curves have the same
asymptotics. Since the causal structure is invariant under conformal
transformations, one can describe this by mapping Minkowski space
to a finite region and adding 'points at infinity'. This is called
a Penrose diagram. In Minkowski space all timelike geodesics
end at the same point,  which is called 'future timelike infinity'.
The backward lightcone of this coint is all of Minkowski space.
If a general space-time contains an asymptotically flat region,
one can likewise introduce a point 
at future timelike infinity. But it might happen
that its backward light cone is not the whole space. In this
case the space-time contains timelike geodesics which do not
'escape' to infinity. The region which is not in the backward light
cone of future timelike infinity is a black hole or a collection
of black holes. The boundary of the
region of no-escape is called a future event horizon. By definition it
is a lightlike surface, i.e. its normal vector field is lightlike.

In Einstein gravity the event horizons of stationary black holes
are so-called Killing horizons.
This property is crucial for the derivation of the zeroth and first
law. A Killing horizon is defined to be a lightlike hypersurface where
a Killing vector field becomes lightlike. For static black holes in
Einstein gravity the horizon Killing vector field is
$\xi = \frac{\der}{\der t}$. Stationary black holes in Einstein
gravity are axisymmetric and the horizon Killing vector 
field is
\be
\xi = \frac{\der}{\der t} + \Omega \frac{\der}{\der \phi} \;,
\ee
where $\Omega$ is the rotation velocity and $\frac{\der}{\der \phi}$
is the Killing vector field of the axial symmetry.

The zeroth and first law do not depend on particular details
of the gravitational field equations. They can be derived in
higher derivative gravity as well, provided one makes the following
assumptions, which in Einstein gravity follow from the field equations:
One has to assume that (i) the event horizon is a Killing horizon and
(ii) that the black hole is either static or that it is stationary,
axisymmetric and posseses a discrete $t-\phi$ reflection symmetry.\footnote{
This means that in adapted coordinates $(t,\phi,\ldots)$
the $g_{t \phi}$-component of the metric vanishes.}

For a Killing horizon one can define the surface gravity $\kappa_S$ by
the equation
\be
\nabla_\m ( \xi^\n \xi_\n) = -2 \kappa_S \xi_\m \;,
\label{DefSFG}
\ee
which is valid on the horizon. 
The meaning of this equation is as follows: The Killing horizon
is defined by the equation $\xi^\n \xi_\n = 0$. The gradient
of the defining equation of a surface is a normal vector field to
the surface. Since $\xi_\m$ is also a normal vector field both have
to be proportional. The factor between the two vectors fields
defines the surface gravity by (\ref{DefSFG}). 
A priori the surface gravity is a 
function on the horizon. But the according to the 
zeroth law of black hole mechanics
it is actually a constant,
\be 
\kappa_S = \mbox{const.}
\ee

The first law of black hole mechanics is energy conservation:
when comparing two infinitesimally close stationary black holes
in Einstein gravity one finds:
\be
\delta M = \frac{1}{8\pi} \kappa_S \delta A + \Omega \delta J
+ \mu \delta Q \;.
\ee
Here 
$A$ denotes the area of the event horizon, $J$ is the 
angular momentum and $Q$ the charge. 
$\Omega$ is the rotation velocity and $\mu = \ft{Q r_+}{A}$.

The comparison of the zeroth and first law of black hole mechanics
to the zeroth and first law thermodynamics,
\be
T = \mbox{const} \;,
\ee
\be
\delta E = T \delta S + p dV + \mu d N \;,
\ee
suggests to identify surface gravity with temperature and 
the area of the event horizon with entropy:
\be
\kappa_S \sim T \;,\;\;\; A \sim S \;.
\ee
Classically this identification does not seem to have physical 
content, because a black hole cannot emit radiation and therefore
has temperature zero. This changes when quantum mechanics is taken
into account: A stationary black hole emits Hawking radiation, which
is found to be proportional to its surface gravity:
\be
T_H = \frac{\kappa_S}{2 \pi} \;.
\ee
This fixes the factor between area and entropy:
\be
S_{BH} = \frac{A}{4} \frac{1}{G_N} \;.
\label{SBH}
\ee
In this formula we reintroduced Newton's constant in order to show
that the black hole entropy is indeed dimensionless (we have set
the Boltzmann constant to unity).
The relation (\ref{SBH}) is known as the area law and 
$S_{BH}$ is called the Bekenstein-Hawking entropy. The Hawking
effect shows that it makes sense to identify $\kappa_S$ with the
temperature, but can we show directly that $S_{BH}$ is the entropy?
And where does the entropy of a black hole come from?

We are used to think about entropy in terms of statistical mechanics.
Systems with a large number of degrees of freedom are conveniently
described using two levels of description: A microscopic description
where one uses all degrees of freedom and a coarse-grained,
macroscopic description where one uses a few observables which
characterize the interesting properties of the system. In the
case of black holes we know a macroscopic description in terms 
of classical gravity. The macroscopic observables are 
the mass $M$, the angular momentum $J$ 
and the charge $Q$, whereas the Bekenstein-Hawking entropy plays
the role of the thermodynamic entropy. What is lacking so far
is a microscopic level of description. For certain extreme
black holes we will discuss a proposal of such a desription in
terms of $D$-branes later. Assuming that we have a microscopic description
the microscopic or statistical entropy is
\be
S_{\mscr{stat}} = \log N(M,Q,J) \;,
\ee
where $N(M,Q,J)$ is the number of microstates which belong to the
same macrostate.
If the interpretation of $S_{BH}$ as entropy is correct, 
then the macroscopic and microscopic entropies must coincide:
\be 
S_{BH} = S_{\mscr{stat}} \;.
\ee
We will see later that this is indeed true for the $D$-brane picture of
black holes.

\subsection{Literature}

Our discussion of gravity and black holes and most of the exercises
follow the book by Wald \cite{Wald:1984}, which we recommend
for further study. The two monographies \cite{Cha:1992} and 
\cite{FroNov:1998} cover various aspects of black hole physics
in great detail.

\section{Black holes in supergravity}
\setcounter{equation}{0}

We now turn to the discussion of black holes in the supersymmetric
extension of gravity, called supergravity. The reason for this is
two-fold. The first is that we want to discuss black holes in
the context of superstring theory, which has supergravity as its low energy
limit. The second reason is that extreme black holes are supersymmetric
solitons. As a consequence quantum corrections are highly constrained
and this can be used to make quantitative tests of the microscopic
$D$-brane picture of black holes.

\subsection{The extreme Reissner-Nordstrom black hole}

Before discussing supersymmetry we will collect several special
properties of extreme Reissner-Nordstrom black holes. These will 
be explained in terms of supersymmetry later.

The metric of the extreme Reissner-Nordstrom black hole is
\be
ds^2 = - \left( 1 - \frac{M}{r} \right)^2 dt^2
+ \left( 1 - \frac{M}{r} \right)^{-2} dr^2 + r^2 d \Omega^2 \;,
\label{extreme}
\ee
where $M=\sqrt{q^2 + p^2}$. By a coordinate transformation
one can make the spatial part of the metric conformally
flat. Such coordinates are called isotropic:
\be
ds^2 = - \left( 1 + \frac{M}{r} \right)^{-2} dt^2
+ \left( 1 + \frac{M}{r} \right)^2 ( dr^2 + r^2 d \Omega^2) \;.
\label{extremeisotropic}
\ee
Note that the new coordinates only cover the region outside the horizon,
which now is located at $r=0$.

The isotropic form of the metric is useful for exploring its special
properties. In the near horizon limit $r \rightarrow 0$ we find
\be
ds^2 = - \frac{r^2}{M^2} dt^2 + \frac{M^2}{r^2} dr^2 + M^2 d \Omega^2 \;.
\label{nearhorizon}
\ee
The metric factorizes asymptotically into two two-dimensional
spaces, which are parametrized by $(t,r)$ and $(\theta,\phi)$, respectively.
The $(\theta,\phi)$-space is obviously a two-sphere of radius $M$,
whereas the $(t,r)$-space is the two-dimensional Anti-de Sitter
space $AdS^2$, with radius $M$. Both are maximally symmetric spaces:
\be
S^2 = \frac{SO(3)}{SO(2)}, \;\;\;
AdS^2 = \frac{SO(2,1)}{SO(1,1)} \;.
\ee
The scalar curvatures of the two factors are proportional to
$\pm M^{-1}$ and precisely cancel, as they must, because 
the product space has a vanishing curvature scalar, $R=0$, 
as a consequence of $T^\m_\m = 0$.

The $AdS^2 \times S^2$ space is known as the Bertotti-Robinson
solution. More precisely it is one particular specimen of the family
of Bertotti-Robinson solutions, which are solutions of Einstein-Maxwell
theory with covariantly constant electromagnetic field strength.
The particular solution found here corresponds to the case with
vanishing cosmological constant and absence of charged matter.

The metric (\ref{nearhorizon}) has one more special property: it is
conformally flat.

\begin{Exercise}
Find the coordinate transformation that maps (\ref{extreme})
to (\ref{extremeisotropic}). Show that in isotropic coordinates
the 'point'
$r=0$ is a sphere of radius $M$ and area $A=4\pi M^2$.
Show that the metric (\ref{nearhorizon}) is conformally flat.
(Hint: It is not necessary to compute the Weyl curvature tensor.
Instead, there is a simple coordinate transformation which makes
conformal flatness manifest.)
\end{Exercise}

We next discuss another astonishing property of the 
extreme Reissner-Nordstrom solution. Let us 
drop spherical symmetry and look for solutions of Einstein-Maxwell
theory with a metric of the form
\be
ds^2 = - e^{-2 f(\vec{x})} dt^2 + e^{2 f(\vec{x})} d \vec{x}^2 \;.
\label{MPmetric}
\ee
In such a background the Maxwell equations are solved 
by electrostatic fields with a potential given in terms of $f(\vec{x})$:
\be
F_{ti} = \mp \der_i ( e^{-f} ) \;,\;\;\;
F_{ij} = 0 \;.
\label{electrostatic}
\ee
More general dyonic solutions which carry both electric and magnetic
charge can be generated by duality transformations. The only
constraint that the coupled Einstein and Maxwell equations impose
on $f$ is that $e^f$ must be a harmonic function,
\be
\Delta e^f = \sum_{i=1}^3 \der_i \der_i e^f = 0 \;.
\label{harmonic}
\ee
Note that $\Delta$ is the flat Laplacian. The solution
(\ref{MPmetric},\ref{electrostatic},\ref{harmonic}) is known as
the Majumdar-Papapetrou solution.

\begin{Exercise}
Show that (\ref{electrostatic}) solves the Maxwell equations in 
the metric background (\ref{MPmetric}) if and only if $e^f$ is 
harmonic.
\end{Exercise}

One possible choice of the harmonic function is
\be
e^f = 1 + \frac{M}{r} \;.
\ee
This so-called single-center solution is the extreme 
Reissner-Nordstrom black hole with mass $M=\sqrt{q^2 + p^2}$.

The more general harmonic function
\be 
e^f = 1 + \sum_{I=1}^{N} \frac{M_I}{ | \vec{x} - \vec{x}_I | }
\ee
is a so-called multi-center solution, which describes
a static configuration of extreme Reissner-Nordstrom black holes with
horizons located at positions $\vec{x}_I$. 

These positions are completely 
arbitrary: gravitational attraction and electrostatic and magnetostatic
repulsion cancel for every choice of $\vec{x}_I$. This is called the 
no-force property. 

The masses of the black holes are
\be
M_I = \sqrt{q_I^2 + p_I^2} \;,
\ee
where $q_I,p_I$ are the electric and magnetic charges.
For purely electric solutions, $p_I=0$, the Maxwell equations
imply that $\pm q_I=M_I$, depending on the choice of
sign in (\ref{electrostatic}).
In order to avoid naked singularities we have
to take all the masses to be positive. As a consequence either all the charges
$q_I$ are positive or they are negative. This is natural, because one needs
to cancel the gravitational attraction by electrostatic repulsion in
order to have a static solution. In the case of a dyonic solution all the
complex numbers $q_I + i p_I$ must have the same phase in the complex 
plane.

Finally one might ask whether other choices of the harmonic function
yield interesting solutions. The answer is no, because all other
choices lead to naked singularities.

Let us then collect the special properties of the extreme 
Reissner-Nordstrom black hole: It saturates the mass bound
for the presence of an event horizon and has vanishing surface
gravity and therefore vanishing Hawking temperature. The
solution interpolates between two maximally symmetric geometries:
Flat space at infinity and the Bertotti-Robinson solution at
the horizon. Finally there exist static multi-center solutions with
the remarkable no-force property.

As usual in physics special properties are expected to be manifestations
of a symmetry. We will now explain that the symmetry 
is (extended) supersymmetry. Moreover the interpolation property
and the no-force property are reminiscent of the Prasad Sommerfield limit
of 't Hooft Polyakov
monopoles in Yang-Mills theory. This is not a  coincidence: 
The extreme Reissner-Nordstrom is a supersymmetric
soliton of extended supergravity.

\subsection{Extended supersymmetry}

We will now review the supersymmetry algebra and its representations.
Supersymmetric theories are theories with conserved spinorial currents.
If $N$ such currents are present, one gets $4N$ real conserved
charges, which can either be organized into $N$ Majorana spinors
$Q^A_m$ or into $N$ Weyl spinors $Q^A_{\alpha}$. Here
$A=1,\ldots, N$ counts the supersymmetries, whereas 
$m=1, \ldots, 4$ is a Majorana spinor index and $\alpha = 1,2$ is a
Weyl spinor index. The hermitean conjugate of $Q^A_{\alpha}$
is denoted by $Q^{+A}_{\alpha}$. It has opposite chirality, but
we refrain from using dotted indices.

According to the theorem of Haag, Lopuzanski and Sohnius the most
general supersymmetry algebra (in four space-time dimensions) is
\bea
\{ Q^A_{\alpha} , Q^{+B}_{\beta} \} &=& 2 \sigma^\m_{\alpha \beta} P_\m 
\d^{AB} \;, \\
\{ Q^A_{\alpha} , Q^{B}_{\beta} \} &=& 2 \ve_{\alpha \beta}
Z^{AB} \;. \\
\nonumber
\eea
In the case of extended supersymmetry, $N>1$, not only the momentum
operator $P_\m$, but also the operators $Z^{AB}$ occur on the 
right hand side of the anticommutation relations. The matrix
$Z^{AB}$ is antisymmetric. The operators in $Z^{AB}$ commute
with all operators in the super Poincar\'e algebra and therefore they are
called central charges. In the absence of central charges the automorphism
group of the algebra is $U(N)$. If central charges are present the 
automorphism group is reduced to 
$USp(2N) = U(N) \cap Sp(2N,\mathbb{C})$.\footnote{Our convention 
concerning the symplectic group is that $Sp(2)$ has rank 1. In other
words the argument is always even.} One can then use $U(N)$ transformations
which are not symplectic to skew-diagonalize the antisymmetric matrix
$Z^{AB}$.

For concreteness we now especialize to the case $N=2$.
We want to construct representations and we start with massive
representations, $M^2 >0$. Then the momentum operator can be brought
to the standard form $P_\m = ( -M , \vec{0} )$. Plugging this into
the algebra and setting $2 |Z| = |Z^{12}|$ the algebra
takes the form
\bea
\{ Q^A_{\alpha} , Q^{+B}_{\beta} \} &=& 2 M \delta_{\alpha \beta}
\d^{AB} \;, \nonumber \\
\{ Q^A_{\alpha} , Q^{B}_{\beta} \} &=& 2 |Z|  \ve_{\alpha \beta}
\ve^{AB} \;.\\
\nonumber
\eea
The next step is to rewrite the algebra using fermionic
creation and annihilation operators. By taking appropriate
linear combinations of the supersymmetry charges one can bring
the algebra to the form
\bea
\{ a_{\alpha}, a^+_{\beta} \} &=& 2 ( M + |Z| ) \d_{\alpha \beta}\;, 
\nonumber\\
\{ b_{\alpha}, b^+_{\beta} \} &=& 2 ( M - |Z| ) \d_{\alpha \beta} \;.
\label{abAlgebra} \\
\nonumber
\eea
Now one can choose any irreducible representation $[s]$ of the little group
$SO(3)$ of massive particles and take the $a_{\alpha},b_{\beta}$ to
be annihilation operators,
\be
a_{\alpha} | s \ra = 0, \;\;\; b_{\beta} | s \ra = 0 \;.
\ee
Then the basis of the corresponding irreducible representation
of the super Poincar\'e algebra is
\be
{\cal B} = \{ a_{\alpha_1}^+ \cdots b_{\beta_1}^+ \cdots | s \ra \} \;.
\ee
In the context of quantum mechanics we are only interested in 
unitary representations. Therefore we have to require the absence
of negative norm states. This implies that the mass is bounded by
the central charge:
\be
M \geq | Z | \;.
\ee
This is called the BPS-bound, a term originally coined in the context
of monopoles in Yang-Mills theory. The representations fall into two
classes. If $M > |Z|$, then we immediately get unitary representations.
Since we have 4 creation operators the dimension is
$2^4 \cdot \dim [s]$. These are the so-called long representations.
The most simple example is the long vector multplet with
spin content $(1 [1], 4 [ \ft12 ], 5 [0] )$. It has 8 bosonic and
8 fermionic on-shell degrees of freedom.

If the BPS bound is saturated, $M=|Z|$, then the representation
contains null states, which have to be devided out in order to get
a unitary representation. This amounts to setting the $b$-operators
to zero. As a consequence half of the supertransformations act trivially.
This is usually phrased as: The multiplet is invariant under half of the
supertransformations.
The basis of the unitary representation is
\be
{\cal B}' = \{ a_{\alpha_1}^+ \cdots | s \ra \} \;.
\ee
Since there are only two creation operators, the dimension is
$2^2 \cdot \dim [s]$. These are the so-called short representations or
BPS representations. Note that the relation $M=|Z|$ is a consequence
of the supersymmetry algebra and therefore cannot be spoiled 
by quantum corrections (assuming that the full theory is supersymmetric).

There are two important examples of short multiplets. One is the short
vector multiplet, with spin content
$( 1 [1], 2 [ \ft12], 1 [0] )$, the other is the hypermultiplet with
spin content $(2 [\ft12], 4 [0] )$. Both have four bosonic and four
fermionic on-shell degrees of freedom.

Let us also briefly discuss massless representations. In this 
case the momentum operator can be brought to the standard form
$P_\m = (-E, 0,0, E)$ and the little group is $ISO(2)$, the two-dimensional
Euclidean group. Irreducible representations of the Poincar\'e group
are labeled by their helicity $h$, which is the quantum number of
the representation of the subgroup $SO(2) \subset ISO(2)$. Similar to
short representations one has to set half of the operators to zero in
order to obtain unitary representations. Irreducible representations of
the super Poincar\'e group are obtained by acting with the remaining
two creation operators on a helicity eigenstate $|h\ra$. Note that
the resulting multiplets will in general not be CP selfconjugate.
Thus one has to add the CP conjugated multiplet to describe the
corresponding antiparticles. There are three important examples of
massless $N=2$ multiplets. The first is the supergravity multiplet
with helicity content
$(1 [ \pm 2], 2 [\pm \ft32], 1 [ \pm 1 ] )$. The states correspond
to the graviton, two real gravitini and a gauge boson, called the
graviphoton. The bosonic field content is precisely the one
of Einstein-Maxwell theory. Therefore Einstein-Maxwell theory can
be embedded into $N=2$ supergravity by adding two gravitini.
The other two important examples of massless multiplets are the
massless vector and hypermultiplet, which are massless versions
of the corresponding massive short multiplets.

In supersymmetric field theories the supersymmetry algebra is realized
as a symmetry acting on the underlying fields. The operator generating an 
infinitesimal  supertransformation takes the form 
$\delta^Q_\epsilon = \overline{\epsilon}_A^m Q_m^A$, when using Majorana
spinors. The transformation paramaters $\epsilon_A^m$ are $N$
anticommuting Majorana spinors. Depending on whether they are constant
or space-time dependent, supersymmetry is realized as a rigid or local
symmetry, respectively. In the local case, the anticommutator of 
two supertransformations yields a local translation, i.e. a general
coordinate transformation. Therefore locally supersymmetric field theories
have to be coupled to a supersymmetric extension of gravity, called
supergravity. The gauge fields of general coordinate transformations and
of local supertransformations are the graviton, described by the 
vielbein $e_\m^{\;\;a}$ and the gravitini $\psi_\m^A = \psi_{\mu m}^A$.
They sit in the supergravity multiplet. We have specified the $N=2$
supergravity multiplet above.

We will now explain why we call the extreme Reissner-Nordstrom 
black hole
a 'supersymmetric soliton'. Solitons are stationary, regular and stable
finite energy solutions to the equations of motion. 
The extreme Reissner-Nordstrom black hole is stationary (even static)
and has finite energy (mass). It is regular in the sense of not having
a naked singularity. We will argue below that it is stable, at least when
considered as a solution of $N=2$ supergravity. What do we mean by
a 'supersymmetric' soliton? Generic solutions to the equations of motion
will not preserve any of the symmetries of the vacuum. In the context
of gravity space-time symmetries are generated by Killing vectors. 
The trivial vacuum, Minkowski space, has ten Killing vectors, because
it is Poincar\'e invariant. A generic space-time will not have
any Killing vectors, whereas special, more symmetric space-times
have some Killing vectors, but not the maximal number of 10. For example
the Reissner-Nordstrom black hole has one timelike Killing vector field
corresponding to time translation invariance and three spacelike
Killing vector fields corresponding to rotation invariance. 
But the spatial translation invariance is broken, as it must be for
a finite energy field configuration. Since the underlying theory is
translation invariant, all black hole solutions related by rigid
translations are equivalent and have in particular the same energy. 
In this way every symmetry of the vacuum which is broken by the field
configuration gives rise to a collective mode.

Similarly a solution is called supersymmetric if it is invariant under
a rigid supertransformation. In the context of locally supersymmetric
theories 
such residual rigid supersymmetries are the fermionic
analogues of isometries.
A field configuration $\Phi_0$ is 
supersymmetric if there exists a choice $\epsilon(x)$ of the
supersymmetry transformation parameters such that the configuration
is invariant,
\be
\left. \delta_{\epsilon(x)} \Phi \right|_{\Phi_0} = 0 \;.
\label{KillingSpinorEq}
\ee
As indicated by notation one has to perform a supersymmetry variation
of all the fields $\Phi$, with parameter $\epsilon(x)$ and then
to evaluate it on the field configuration $\Phi_0$. The transformation
parameters $\epsilon(x)$ are fermionic analogues of Killing vectors
and therefore they are called Killing spinors. Equation
(\ref{KillingSpinorEq}) is referred to as the Killing spinor equation.
As a consequence of the residual supersymmetry the number of 
fermionic collective modes is reduced. If the solution is particle like,
i.e. asymptotically flat and of finite mass, then we expect that it
sits in a short multiplet and describes a BPS state of the theory.

Let us now come back to the extreme Reissner-Nordstrom black hole.
This is a solution of Einstein-Maxwell theory, which can be embedded
into $N=2$ supergravity by adding two gravitini $\psi^A_\m$.
The extreme Reissner-Nordstrom black hole is also a solution
of the extended theory, with $\psi^A_\m=0$. Moreover it is a supersymmetric
solution in the above sense, i.e. it posesses Killing spinors.
What are the Killing spinor equations in this case? 
The graviton $e_\m^{\;\;a}$ and
the graviphoton $A_\m$ transform into fermionic quantities, which all
vanish when evaluated in the background.
Hence the only conditions come from the 
gravitino variation:
\be
\delta_{\epsilon} \psi_{\m A}
= \nabla_\m \epsilon_A - \ft14 F^-_{ab} \gamma^a \gamma^b
\gamma_\m \ve_{AB} \epsilon^B \stackrel{!}{=} 0 \;.
\label{gravitinovariation}
\ee
The notation and conventions used in this equation are as follows:
We suppress all spinor indices and use the so-called chiral notation.
This means that we use four-component Majorana spinors, but project onto
one chirality, which is encoded in the position of the supersymmetry 
index $A=1,2$:
\be
\gamma_5 \epsilon^A = \epsilon^A \;\;\; \gamma_5 \epsilon_A = 
- \epsilon_A \;.
\ee
As a consequence of the Majorana condition only half of the
components of $\epsilon_A, \epsilon^A$ are independent, i.e. there are
8 real supertransformation parameters. The indices $\m,\n$ are
curved and the indices $a,b$ are flat tensor indices. $F_{\m\n}$ is
the graviphoton field strength and 
\be
F^{\pm}_{\m\n} = \ft12 \left( F_{\m \n} \pm i \; {}^{\star} F_{\m \n} \right)
\ee
are its selfdual and antiselfdual part.

One can now check that the Majumdar-Papapetrou solution 
and in particular the extreme Reissner-Nordstrom black hole
have Killing spinors
\be
\epsilon_A (\vec{x}) = h(\vec{x}) \epsilon_A(\infty)\;,
\ee
where $h(\vec{x})$ is completely fixed in terms of $f(\vec{x})$.
The values of the Killing spinors at infinity are 
subject to the condition
\be
\epsilon_A(\infty) + i \g^0 \frac{Z}{|Z|} \ve_{AB} \epsilon^B(\infty) = 0 \;.
\ee
This projection fixes half of the parameters in terms of the others. 
As a consequence we have four Killing spinors, which is half of the 
maximal number eight. The four supertransformations which do not act
trivially
correspond to four fermionic collective modes. It can be shown
that the extreme Reissner-Nordstrom black hole is part of a hypermultiplet.
The quantity $Z$ appearing in the phase factor $Z/|Z|$ is the central charge.
In locally supersymmetric theories the central charge transformations
are local $U(1)$ transformations, and the corresponding gauge field 
is the graviphoton. The central charge is a complex linear combination of
the electric and magnetic charge of this $U(1)$:
\be
Z = \frac{1}{4 \pi} \oint 2 F^- = p-iq \;.
\label{CentralChargeLocal}
\ee
Since the mass of the extreme Reissner-Nordstrom black hole
is $M = \sqrt{q^2 + p^2} = |Z|$ we see that the extreme limit
coincides with the supersymmetric BPS limit. The extreme 
Reissner-Nordstrom black hole therefore has all the properties
expected for a BPS state: It is invariant under half of the 
supertransformations, sits in a short multiplet and saturates the
supersymmetric mass bound. We therefore expect that it 
is absolutely stable, as a solution
of $N=2$ supergravity. Since the surface gravity and, hence, the
Hawking temperature vanishes it is stable against Hawking radiation.
It is very likely, however,  that charged black holes in non-supersymmetric
gravity are unstable due to charge superradiance. But in a theory 
with $N\geq 2$ supersymmetry there is no state of lower energy and
the black hole is absolutely stable. Note also that the no-force property
of multi-center solutions can now be understood as a consequence of
the additional supersymmetry present in the system.

Finally we would like to point out that supersymmetry also accounts
for the special properties of the near horizon solution. Whereas
the BPS black hole has four Killing spinors at generic values of the
radius $r$, this
is different at infinity and at the horizon. At infinity the solution
approaches flat space, which has 8 Killing spinors. But also 
the Bertotti-Robinson geometry, which is  approached
at the horizon, has 8 Killing spinors.
Thus the number of unbroken supersymmetries doubles in the asymptotic
regions. Since the Bertotti-Robinson solution has the maximal number
of Killing spinors, it is a supersymmetry singlet and 
an alternative vacuum of $N=2$ supergravity. Thus, the extreme
Reissner-Nordstrom black hole interpolates between vacua: this is
another typical property of a soliton.

So far we have seen that one can check that a given solution to the 
equations of motion is
supersymmetric, by plugging it into the Killing spinor equation.
Very often one can successfully proceed in the opposite way and
systematically construct supersymmetric solutions by first looking at
the Killing spinor equation and taking it as a condition on the
bosonic background. This way one gets first order differential equations
for the background which are more easily solved then the equations of
motion themselves, which are second order. Let us illustrate this with 
an example.

\begin{Exercise}
Consider a metric of the form
\be
ds^2 = - e^{-2 f(\vec{x}) } dt^2 + e^{ 2 f(\vec{x})} d\vec{x}^2 \;,
\ee
with an arbitrary function $f(\vec{x})$. In such a background the
time component of the Killing spinor equation takes the form
\be
\delta \psi_{tA} = - \frac{1}{2} \der_i f e^{-2f} \gamma^0 \gamma^i 
\epsilon_A + e^{-f} F_{0i}^- \g^i \ve_{AB} \epsilon^B \stackrel{!}{=}0\;.
\ee
In comparison to (\ref{gravitinovariation}) we have chosen the
time component and explicitly evaluated the spin connection.
The indices $0,i=1,2,3$ are flat indices.

Reduce this equation to one differential equation for the background
by making an ansatz for the Killing spinor. Show that the resulting
equation together with the Maxwell equation for the graviphoton
field strength implies that this solution is precisely the Majumdar
Papapetrou solution.
\end{Exercise}

As this exercise illustrates, the problem of constructing supersymmetric
solutions has two parts. The first question is what algebraic condition
one has to impose on the Killing spinor. This is also the most important
step in classifying supersymmetric solitons.
In a second step one has to determine the bosonic background
by solving differential equations. As illustrated in the above exercise
the resulting solutions are very often expressed in terms of harmonic
functions. We would now like to discuss the first, algebraic step
of the problem. This is related to the so-called Nester construction.
In order to appreciate the power of this formalism we digress for a 
moment from our main line of thought and discuss positivity theorems
in gravity.

Killing spinors are useful even outside supersymmetric theories.
The reason is that one can use the embedding of a non-supersymmetric
theory into a bigger supersymmetric
theory as a mere tool to derive results. One famous example is the
derivation of the positivity theorem for the ADM mass of asymptotically
flat space-times by Witten, which, thanks to the use of spinor techniques
is much simpler then the original proof by Schoen and Yau. The Nester
construction elaborates on this idea.

In order to prove the positivity theorem one makes certain general
assumptions: One considers an asymptotically flat space-time, 
the equations of motion are required to be satisfied and it is assumed
that the behaviour of matter is 'reasonable' in the sense that a
suitable condition on the energy momentum tensor (e.g. the so-called
dominant energy condition) is satisfied.
The Nester construction then tells how to construct
a two-form $\omega_2$, such that the integral
over an asymptotic two-sphere satisfies the inequality
\be
\oint \omega_2 = \overline{\epsilon(\infty)} [ \gamma^\m P_\m + ip 
+ q \gamma_5 ] \epsilon(\infty) \geq 0 \;.
\ee
Here $P_\m$ is the four-momentum of the space-time (which is defined
because we assume asymptotic flatness), $q,p$ are its electric and
magnetic charge and $\epsilon(\infty)$ is the asymptotic value
of a spinor field used as part of the construction. (The spinor
is a Dirac spinor.) The matrix between the spinors is 
called the Bogomol'nyi matrix (borrowing again terminology from 
Yang-Mills theory). It has eigenvalues
$M \pm \sqrt{q^2 + p^2}$ and therefore we get prescisely
the mass bound familiar from the Reissner-Nordstrom black hole. 
But note that this result has been derived based on general assumptions,
not on a particular solution.
Equality holds if and only if the spinor field
$\epsilon(x)$ satisfies the Killing spinor equation 
(\ref{gravitinovariation}). The static space-times satisfying the bound are
prescisely the Majumdar-Papapetrou solutions.

The relation to supersymmetry is obvious: we have seen above that
the matrix of supersymmetry anticommutators has eigenvalues
$M \pm |Z|$, (\ref{abAlgebra}) and that in supergravity the central charge is
$Z = p - iq$, (\ref{CentralChargeLocal}). 
Thus the Bogomonl'nyi matrix must be related to the matrix
of supersymmetry anticommutators.

\begin{Exercise}
Express the Bogomol'nyi matrix in terms of supersymmetry 
anticommutators.
\end{Exercise}

The algebraic problem of finding the possible projections of
Killing spinors is equivalent to finding the possible eigenvectors
with eigenvalue zero of the Bogomol'nyi matrix. Again we will
study one particular example in an exercise.

\begin{Exercise}
Find a zero eigenvector of the Bogomonl'nyi matrix which desrcibes
a massive BPS state at rest.
\end{Exercise}

In the case of pure $N=2$ supergravity all supersymmetric solutions
are known.
Besides the Majumdar-Papapetrou solutions there are two further 
classes of solutions:
The Israel-Wilson-Perjes (IWP) solutions, which are rotating, stationary
generalizations of the Majumdar-Papapetrou solutions and the 
plane fronted gravitational waves with parallel rays (pp-waves).

\subsection{Literature}

The representation theory of the extended supersymmetry algebra
is treated in chapter 2 of Wess and Bagger
\cite{WesBag:1992}.
The interpretation of the extreme Reissner-Nordstrom black hole
as a supersymmetric soliton is due to Gibbons \cite{Gib:1981}.
Then Gibbons and Hull showed that the Majumdar-Papetrou solutions
and pp-waves are supersymmetric \cite{GibHul:1982}. They also discuss
the relation to the positivity theorem for the ADM mass and the Nester
construction. The classification of supersymmetric solitons 
in pure $N=2$ supergravity was completed by Tod \cite{Tod:1983}.
The Majumdar-Papetrou solutions are discussed in some detail 
in \cite{Cha:1992}. Our discussion of Killing spinors uses 
the conventions of Behrndt, L\"ust and Sabra, who have treated the
more general case where vector multiplets are coupled to 
$N=2$ supergravity \cite{BehLueSab:1997}. A nice exposition
of how supersymmetric solitons are classified in terms of 
zero eigenvectors of the Bogomol'nyi matrix has been given
by Townsend for the case of eleven-dimensional supergravity 
\cite{Tow:1997}.

\section{$p$-branes in type II string theory}
\setcounter{equation}{0}

In this section we will consider $p$-branes, which are higher
dimensional cousins of the extremal Reissner-Nordstrom black hole.
These $p$-branes are supersymmetric solutions of ten-dimensional
supergravity, which is the low energy limit of string theory. We will
restrict ourselves to the string theories with the highest possible
amount of supersymmetry, called type IIA and IIB. 
We start by reviewing the relevant elements of string theory.

\subsection{Some elements of string theory}

The motion of a string in a curved space-time background
with metric $G_{\m \n}(X)$ is described by a two-dimensional
non-linear sigma-model with action
\be
S_{WS} = \frac{1}{4 \pi \alpha'} \int_{\Sigma} 
d^2 \sigma \sqrt{-h} h^{\alpha
\beta}(\sigma) \der_{\alpha} X^{\m} \der_{\beta} X^\n G_{\m\n}(X) \;.
\label{WSaction}
\ee
The coordinates on the world-sheet $\Sigma$ 
are $\sigma=(\sigma^0,\sigma^1)$
and $h_{\alpha \beta}(\sigma)$ is the intrinsic world-sheet metric,
which locally can be brought to the flat form $\eta_{\alpha \beta}$.
The coordinates of the string in space-time are $X^\m(\sigma)$.
The parameter $\alpha'$ has the dimension $L^2$ (length-squared) and is
related to the string tension $\tau_{F1}$ by 
$\tau_{F1} = \ft1{2 \pi \alpha'}$. It is the only
independent dimensionful parameter in string theory. Usually one
uses string units, where $\alpha'$ is set to a constant (in addition
to $c=\hbar=1$).\footnote{We will see later that it is in general
not possible to use Planckian and stringy units simultantously.
The reason is that the ratio of the Planck and string scale is
the dimensionless string coupling, which is related to the vacuum
expectation value of the dilaton and which is a free parameter,
at least in perturbation theory.}
In the case of a flat space-time background,
$G_{\m \n} = \eta_{\m \n}$, the world-sheet action (\ref{WSaction})
reduces to the action of $D$ free two-dimensional scalars and the 
theory can be quantized exactly. In particular one can identify
the quantum states of the string.

At this point one can define different theories by specifying 
the types of world sheets that one admits. Both orientable and 
non-orientable world-sheets are possible, but we will only
consider orientable ones. Next one has the freedom of adding
world-sheet fermions. Though we are interested in type II
superstrings, we will for simplicity first consider bosonic
strings, where no world-sheet fermions are present. 
Finally one has to specify the boundary 
conditions along the space direction of the world sheet. 
One choice is to impose Neumann boundary conditions,
\be
\left. \der_1 X^\m  \right|_{\der \Sigma} = 0 \;.
\ee
This corresponds to open strings. In the following we will be
mainly interested in the massless modes of the strings, because
the scale of massive excitations is naturally of the order of
the Planck scale. The massless state of the open bosonic string
is a gauge boson $A_\m$.

Another choice of boundary conditions is Dirichlet boundary conditions,
\be
\left. \der_0 X^\m \right|_{\der \Sigma} = 0 \;.
\ee
In this case the endpoints of the string are fixed. Since momentum
at the end is not conserved, such boundary conditions require to 
couple the string to another dynamical object, called 
a $D$-brane. Therefore Dirichlet
boundary conditions do not describe strings in the vacuum but in
a solitonic background. Obviously the corresponding soliton
is a very exotic object, since we can describe it in a perturbative
picture, whereas conventional solitons are invisible in perturbation
theory.
As we will see later $D$-branes have a complementary 
realization as higher-dimensional 
analogues of extremal black holes. The perturbative 
$D$-brane picture of black holes can be used to count microstates 
and to derive the microscopic entropy.

In order to prepare for this let us consider a situtation where
one imposes Neumann boundary conditions along time and along
$p$ space directions and Dirichlet boundary conditions along
the remaining $D-p-1$ directions ($D$ is the dimension of space-time).
More precisely we require that open strings end on the
$p$-dimensional plane $X^m = X^m_0$, $m=p+1,\ldots,D-p-1$.
This is called a Dirichlet-$p$-brane or $Dp$-brane for short.
The massless states are obtained from the case of pure Neumann
boundary conditions by dimensional reduction:
One gets a $p$-dimensional gauge boson
$A^\m$, $\m = 0,1,\ldots,p$ and $D-p-1$ scalars $\phi^m$.
Geometrically the scalars describe transverse oscillations of the
brane.

As a generalization one can consider $N$ parallel $Dp$-branes.
Each brane carries a $U(1)$ gauge theory on its worldvolume, and
as long as the branes are well separated these are the only light states.
But if the branes are very close, then additional light states
result from strings that start and end on different branes. These
additional states complete the adjoint representation of $U(N)$ and
therefore the light excitations of $N$ near-coincident $Dp$ branes
are described by the dimensional reduction of  $U(N)$ gauge theory
from $D$ to $p+1$ dimensions.

The final important class of boundary conditons are periodic
boundary conditions. They describe closed strings. The massless
states are the graviton $G_{\m\n}$, an antisymmetric tensor 
$B_{\m\n}$ and a scalar $\phi$, called the dilaton. As indicated
by the notation a curved background as in the action (\ref{WSaction})
is a coherent states of graviton string states. One can generalize this
by adding terms which describe the coupling of the string to other classical
background fields. For example the couplings to the $B$-field and to
the open string gauge boson $A_\m$ are
\be
S_B = \frac{1}{4 \pi \alpha'} 
\int_{\Sigma} d^2 \sigma \ve^{\alpha \beta} \der_{\alpha} X^\m
\der_{\beta} X^\n B_{\mu \nu}(X) 
\label{Baction}
\ee
and
\be
S_A = \oint_{\der \Sigma} d^1 \sigma^{\alpha}  \der_{\alpha} X^\m
A_\m(X) \;.
\ee

Interactions of strings are encoded in the topology of the world-sheet. 
The S-matrix can be computed
by a path integral over all world sheets connecting given initial
and final states. For the low energy sector all the 
relevent information is contained in the low energy effective action of the 
massless modes.
We will see examples later. The low energy effective action is derived
by either matching string theory amplitudes with field theory amplitudes
or by imposing that the non-linear sigma model, which describes the
coupling of strings to the background fields $G_{\m \n}, B_{\m\n},\ldots$
is a conformal field theory. Conformal invariance
of the world sheet theory is necessary for keeping the world sheet
metric $h_{\alpha \beta}$ non-dynamical.\footnote{It might be possible 
to relax this and to consider the so-called non-critical or
Liouville string theory. But then one gets a different and much more
complicated theory.} A set of background fields $G_{\mu \nu},
B_{\mu \nu}, \ldots$ which leads to an exact conformal field theory
provides an exact solution to the classical equations of motion of 
string theory. Very often
one only knows solutions of the low energy effective field 
theory.

\begin{Exercise}
Consider a curved string background which is independent
of the coordinate $X^1$, and with $G_{1 \n}=0$, $B_{1 \n}=0$ and
$\phi=\mbox{const}$. Then the $G_{11}$-part of the world-sheet
action factorizes,\footnote{We have set $\alpha'$ to a constant for
convenience.}
\be
S[G_{11}] = \int d^2 \sigma 
G_{11}(X^m) \der_+ X^1 \der_- X^1 \;,
\label{global}
\ee 
where $m \not=1$. We have introduced light-cone coordiantes
$\sigma^{\pm}$ on the world-sheet. The isometry of the target space
$X^1 \rightarrow X^1 + a$, where $a$ is a constant, is a global
symmetry from the world-sheet point of view. Promote this to
a local shift symmetry, $X^1 \rightarrow X^1 + a(\sigma)$
('gauging of the global symmetry') by introducing suitable
covariante derivatives $D_{\pm}$. Show that the locally
invariant action
\be
\hat{S} = \int d^2 \sigma \left( 
G_{11} D_+ X^1 D_- X^1 + \tilde{X}^1 F_{+-}
\right) \;,
\label{local}
\ee
where $F_{+-} = [D_+, D_-]$ reduces to the globally
invariant action  (\ref{global}), when eliminating 
the Lagrange multiplyer $\tilde{X}^1$ through its equation
of motion. Next, eliminate the gauge field $A_{\pm}$ from
(\ref{local}) through
its equation of motion. What is the interpretation of the
resulting action?
\end{Exercise}

The above exercise illustrates T-duality in the most simple
example. T-duality is a stringy symmetry, which identifies
different values of 
the background fields $G_{\m\n},B_{\m\n},\phi$ 
in a non-trivial way. 
The version of 
T-duality that we consider here applies if the space-time
background 
has an isometry or an abelian group of isometries.
This means that when using adapted coordinates the metric
and all other background fields are independent of one
or of several of the embedding coordinates $X^\m$.
For later reference we note the transformation law of the
fields  under a T-duality transformation along the 1-direction: 
\[
G'_{11} = \frac{1}{G_{11}} \;, \;\;\;
G'_{1m} = \frac{B_{1m}}{G_{11}} \;, \;\;\;
B'_{1m} = \frac{G_{1m}}{G_{11}} \;, 
\]
\[
G'_{mn} = G_{mn} - \frac{ G_{m1} G_{1n} + B_{m1} B_{1n} }{G_{11}} \;,
\;\;\;
B'_{mn} = B_{mn} - \frac{G_{m1} B_{1n} + B_{m1} G_{1n}}{G_{11}} \;,
\]
\be
\phi' = \phi - \log \sqrt{G_{11}} \;,
\label{Buscher}
\ee
where $m\not=1$. These formulae apply to closed bosonic strings.
In the case of open strings T-duality mutually exchanges 
Neumann and Dirichlet boundary conditions. This is one of the
motivations for introducing $D$-branes.

Let us now discuss the extension from the bosonic to the type
II string theory. In type II theory the world sheet action is
extended to a $(1,1)$ supersymmetric action by adding world-sheet
fermions $\psi^{\mu}(\sigma)$. It is a theory of closed oriented strings.
For the fermions one can choose the boundary conditions for the
left-moving and right-moving part independently to be either
periodic (Ramond) or antiperiodic (Neveu-Schwarz). This gives
four types of boundary conditions, which are referred to as
NS-NS, NS-R, R-NS and R-R in the following. Since the ground state
of an R-sector carries a representation of the $D$-dimensional 
Clifford algebra, it is a space-time spinor. Therefore the
states in the NS-NS and R-R sector are bosonic, whereas the states
in the NS-R and R-NS sector are fermionic.

Unitarity of the quantum theory imposes consistency conditions
on the theory. First the space-time dimension is fixed to be $D=10$.
Second one has to include all possible choices of the boundary
conditions for the world-sheet fermions. Moreover the relative weights 
of the various sectors in the string path integral are not arbitrary.
Among the possible choices two lead to supersymmetric theories,
known as type IIA and type IIB. Both differ in the relative 
chiralities of the R-groundstates: The IIB theory is chiral, the
IIA theory is not.

The massless spectra of the two theories are as follows:
The NS-NS sector is identical for IIA and IIB:
\be
\mbox{NS-NS}: G_{\m\n} , B_{\m \n}, \phi \;.
\ee
The R-R sector contains various $n$-form gauge fields
\be
A_n = \frac{1}{n!} A_{\m_1 \cdots \m_n} dx^{\m_1} \wedge \cdots
\wedge dx^{\m_n}
\ee
and is different for the two theories:
\be
\mbox{R-R}: \left\{ \begin{array}{ll}
\mbox{IIA}: & A_1, A_3 \\
\mbox{IIB}: & A_0, A_2, A_4 \\
\end{array} \right.
\ee
The 0-form $A_0$ is a scalar with a Peccei-Quinn symmetry, i.e. it
enters the action only via its derivative. The 4-form is constrained,
because the corresponding field strength $F_5$ is required to be
selfdual: $F_5 = {}^{\star}F_5$. Finally, the fermionic sectors
contain two gravitini and two fermions, called dilatini:
\be
\mbox{NS-R/R-NS}: \left\{ \begin{array}{ll}
\mbox{IIA}: & \psi^{(1)\m}_+, \psi^{(2)\m}_-,\psi^{(1)}_+, \psi^{(2)}_-, \\
\mbox{IIB}: & \psi^{(1)\m}_+, \psi^{(2)\m}_+,\psi^{(1)}_+, \psi^{(2)}_+ \;. \\
\end{array} \right.
\ee

More recently it has been proposed to add $D$-branes and the corresponding
open string sectors to the type II theory. The motivation for this
is the existence of $p$-brane solitons in the type II low energy 
effective theory. In the next sections we will study these
$p$-branes in detail and review the arguments that relate them to
$Dp$-branes. One can show that the presence of a $Dp$-brane or of
several parallel $Dp$-branes breaks only half of the ten-dimensional 
supersymmetry of type IIA/B theory, if one chooses $p$ to be
even/odd, respectively. 
Therefore such backgrounds describe BPS states. The massless
states associated with a $Dp$-brane correspond to the dimensional
reduction of a ten-dimensional vector multiplet from ten to $p+1$
dimensions. In the case of $N$ near-coincident $Dp$-branes one
gets the dimensional reduction of a supersymmetric ten-dimensional $U(N)$ 
gauge theory.

T-duality can be extended to type II string theories. There is one
important difference to the bosonic string: T-duality is not
a symmetry of the IIA/B theory, but maps IIA to IIB and vice versa.

This concludes our mini-introduction to string theory. From now on we
will mainly consider the low energy effective action.

\subsection{The low energy effective action}

The low energy effective action of type IIA/B superstring theory
is type IIA/B supergravity. The $p$-branes which we will discuss
later in this section are solitonic solutions of supergravity.
We need to make
some introductory remarks on the supergravity actions. Since
we are interested in bosonic solutions, we will only discuss the
bosonic part. We start with the NS-NS sector, which is the same for type IIA
and type IIB and contains the graviton $G_{\m \n}$, the antisymmetric
tensor $B_{\m\n}$ and the dilaton $\phi$. One way to parametrize the
action is to use the so-called string frame:
\be
S_{NS-NS} = \frac{1}{2\kappa^2_{10}} 
\int d^{10}x \sqrt{ -G} e^{-2 \phi} \left(R
+ 4 \der_\m \phi \der^\m \phi - \frac{1}{12} H_{\m\n\rho}
H^{\m\n\rho} \right) \;.
\label{IINSstringframe}
\ee
The three-form $H = dB$ is the field strength of the $B$-field.
The metric $G_{\m\n}$ is the string frame metric, that is the metric
appearing in the non-linear sigma-model (\ref{WSaction}), 
which describes the motion of
a string in a curved background. The string frame action is adapted
to string perturbation theory, because it depends on the dilaton in
a uniform way. The vacuum expectation value of the dilaton defines
the dimensionless string coupling,
\be
g_S = e^{ \la \phi \ra} \;.
\ee
The terms displayed in (\ref{IINSstringframe}) are of order $g_S^{-2}$
and arise at string tree level. Higher order $g$-loop contribtutions
are of order $g_S^{-2 + 2g}$ and can be computed using string perturbation
theory. The constant $\kappa_{10}$ has the dimension of a ten-dimensional
gravitational coupling. Note, however, that it can not be directly
identified with the physical gravitational coupling, because
a rescaling of $\kappa_{10}$ can be compensated by a shift of the dilaton's
vacuum expectation value 
$\la \phi \ra$. This persists to higher orders in string perturbation
theory, because $\phi$ and $\kappa_{10}$ only appear in the combination
$\kappa_{10} e^{\phi}$. One can use this to eliminate the scale
set by the dimensionful coupling in terms of the string scale 
$\sqrt{\alpha'}$ by imposing 
\be
\kappa_{10} = ( \alpha')^{2 } g_S \cdot \mbox{const}\;.
\label{kappaalpha}
\ee
There is only one independent dimensionful parameter and only one single
theory, which has a family of ground states parametrized by the
string coupling.

It should be noted that (\ref{IINSstringframe}) has not
the canonical form of a gravitational action. In particular the 
first term is not the standard Einstein-Hilbert term. This is the second
reason why 
the constant $\kappa_{10}$ in front of the action is not necessarily
the physical
gravitational coupling. Moreover the definitions of mass and
Bekenstein-Hawking entropy are
tied to a canonically normalized gravitational
action. Therefore we need to know how to bring (\ref{IINSstringframe})
to canonical form by an appropriate field redefinition. For later
use we discuss this for general space-time dimension $D$. Given
the $D$-dimensional version of (\ref{IINSstringframe}) the 
canonical or Einstein metric is
\be
g_{\m\n} = G_{\m\n} e^{ -4 (\phi - \la \phi \ra ) / (D-2)}
\ee
and the Einstein frame action is
\be
S_{NS-NS} = \frac{1}{2 \kappa_{D,phys}^2} \int d^D x \sqrt{-g}
\left( R(g) + \cdots \right) \;.
\ee
We have absorbed the 
dilaton vacuum expectation value 
in the prefactor of the action to get the physical 
gravitational coupling.\footnote{There is a second, slightly
different definition of the Einstein frame where the dilaton
expectation values is absorbed in $g_{\m \n}$ and not in the
gravitational coupling. This second definition is convenient in the
context of IIB S-duality, because the resulting metric
$g_{\m\n}$ is invariant under S-duality.
The version we use in the text is the correct one if one wants
to use the standard formulae of general relativity to compute
mass and entropy.}
The action now has canonical form, but
the uniform dependence on the string coupling is lost.

Let us now turn to the R-R sector, which consists of $n$-form
gauge fields $A_n$, with $n=1,3$ for type A and $n=0,2,4$
for type B. The standard kinetic term for an $n$-form gauge field
in $D$ dimensions is
\be
S \simeq \int_D F_{n+1} \wedge {}^{\star}F_{n+1} \:,
\label{RRaction}
\ee
where the integral is over $D$-dimensional space and $F_{n+1} = d A_n$.
The R-R action in type II theories contains further
terms, in particular Chern-Simons terms. Moreover the
gauge transformations are more complicated than
$A_n \rightarrow A_n + d f_{n-1}$, because some of the $A_n$ 
are not inert under the transformations of the others.
For simplicity we will ignore these complications here and only
discuss simple
$n$-form actions of the type (\ref{RRaction}).\footnote{The full
R-R action discussion is discussed in \cite{Pol:1998}.} 
We need, however, to make
two further remarks. The first is that the action (\ref{RRaction})
is neither in the string nor in the Einstein frame.
Though the Hodge-$\star$ is build using the string metric, there
is no explicit dilaton factor in front. The reason is that if one
makes the dilaton explicit, then the gauge transformation law
involves the dilaton. It is convenient to have the standard gauge
transformation and as a consequence the standard form of the conserved
charge. Therefore the dilaton has been absorbed in the gauge field
$A_n$ in (\ref{RRaction}), although this obscures the fact that the
term arises at string tree level. The second remark concerns the
four-form $A_4$ in type IIB theory. Since the associated field strength
$F_5$ is selfdual, $F_5 = {}^{\star}F_5$, it is non-trivial to 
write down a covariant action. The most simple way to procede is
to use a term $S \simeq \ft12 \int_{10} F_5 \wedge {}^{\star} F_5$
in the action and to impose $F_5={}^{\star}F_5$ at the level of the
field equations.\footnote{In
\cite{PasSorTon:1997} 
a proposal has been made how to construct covariant actions
for this type of theories.}

Let us now discuss what are the analogues of point-like sources
for an action of the type (\ref{RRaction}). In general electric 
sources which couple minimally to the gauge field $A_n$ are described
by a term
\be
\int_D A_n \wedge {}^{\star} j_n \;,
\label{source}
\ee
where the electric current $j_n$ is an $n$-form. Variation of 
the combined action (\ref{RRaction}),(\ref{source}) yields Euler Lagrange
equations with a source term,
\be
d \; {}^{\star} F_{n+1} = {}^{\star} j_{n} \;.
\label{inhomeq}
\ee
The Bianchi identity is $dF_{n+1}=0$. Analogues of point sources are
found by localizing the current on a $(p=n-1)$-dimensional 
spacelike surface with
$(p+1=n)$-dimensional world volume:
\be
\int_D A_{p+1} \wedge {}^{\star} j_{p+1} = \int_{p+1} A_{p+1} \;.
\label{localizedsource}
\ee
Thus sources are $p$-dimensional membranes, or $p$-branes for short.
We consider the most simple example where space-time is flat and
the source is the  $p$-dimensional plane 
$x^i=0$ for $i=p+1, \ldots, D-1$. 
It is convenient to introduce spheric coordinates in the directions
transverse to the brane, $x^i =  (r, \phi^1,\ldots,\phi^{D-p-2})$.
Then the generalized Maxwell equations reduce to
\be
\Delta^{\perp} A_{01\ldots p}(r) \simeq \delta(r) \;,
\label{harmA}
\ee
where $\Delta^{\perp}$ is the Laplace Operator with respect to the
transverse coordinates and the indices $0,1,\ldots,p$  
belong to directions parallel to the world volume. In the following
we will not keep track of the precise factors in the equations.
This is indicated by the symbol $\simeq$.
The gauge field and field strength solving (\ref{harmA}) are
\be
A_{01\ldots p} \simeq \frac{Q}{r^{D-p-3}} \mbox{   and    }
F_{0r 1 \ldots p} \simeq \frac{Q}{r^{D-p-2}} \;.
\label{flatelectric}
\ee
More generally one might consider a curved space-time or
sources which have a finite extension along the transverse 
directions. If the solution has isometry group
$\mathbb{R}_t \times ISO(p) \times SO(D-p-1)$ and approaches
flat space in the transverse directions, then its asymptotic
form is given by (\ref{flatelectric}).

The parameter $Q$ is the electric charge (or more precisely the
electric charge density). As in electrodynamics one can
write the charge as a surface integral,
\be
Q \simeq \oint_{D-p-2} {}^{\star} F_{p+2} \;,
\ee
where the integration is over a $(D-p-2)$-surface which encloses
the source in transverse space. We take this integral as our 
definition of $p$-brane charge.

Magnetic sources are found by exchanging the roles of 
equation of motion and Bianchi identity. They couple minimally
to the magnetic potential $\tilde{A}$, where $d \tilde{A} = {}^{\star}F$.
Localized sources are $\tilde{p}$-branes with $\tilde{p}= D-p-4$.
The potential and field strength corresponding to a flat $\tilde{p}$-brane
in flat space-time are
\be
\tilde{A}_{01\ldots \tilde{p}} \simeq \frac{P}{r^{p+1}}
\mbox{   and   }
F_{\phi^1 \ldots \phi^{p+2}} \simeq
{}^{\star} F_{0r 1 \ldots \tilde{p}} \simeq \frac{P}{r^{p+2}} \;.
\ee
The magnetic charge is
\be
P \simeq \oint_{p+2} F_{p+2} \;.
\ee

Generically, electric and magnetic sources have different dimensions,
$p\not=\tilde{p}$. Dyonic objects are only possible
for special values $D$ and $p$,
for example 0-branes (particles) in $D=4$ and 
3-branes in $D=10$. Electric and magnetic charges are restricted by
a generalized Dirac quantization condition,
\be
P Q \simeq  n  \;, \mbox{  with   } n \in \mathbb{Z}  \;.
\ee
This can be derived by either generalizing the Dirac string construction
or the Wu-Yang construction.

We now turn to the discussion of $p$-brane solitons in type II 
supergravity. By $p$-brane we indicate that we require that the
soliton has isometries $\mathbb{R}_t \times ISO(p) \times SO(D-p-1)$.
As before we take a soliton to be a solution to the equations of motion,
which has finite energy per worldvolume, 
has no naked singularities and is stable. As in the Reissner-Nordstrom
case one can find solutions which have 
Killing spinors and therefore are stable as a consequence
of the BPS bound.\footnote{For these extremal $p$-branes the isometry group
is enhanced to $ISO(1,p) \times SO(D-p-1)$.}  
The $p$-branes are charged with respect to the
various tensor fields appearing in the type IIA/B action. Since we know
which tensor fields exist in the IIA/B theory, 
we know in advance which solutions we have to expect.
The electric and magnetic source for the $B$-field are a 1-brane
or string, called the fundamental string and a 5-brane, called the
solitonic 5-brane or NS-5-brane. In the R-R-sector there are
R-R-charged $p$-branes with $p=0,\ldots,6$ with $p$ even/odd for
type IIA/B.
Before discussing them we comment on some exotic objects,
which we won't discuss further. First there are R-R-charged $(-1)$-branes
and 7-branes, which are electric and magnetic sources for the type IIB
R-R-scalar $A_0$. The $(-1)$-brane is localized in space and time and
therefore it is interpreted, after going to Euclidean time, as
an instanton. The 7-brane is also special, because it is not
asymptotically flat. This is a typical feature of brane solutions
with less than 3 transverse directions, for example black holes
in $D=3$ and cosmic strings in $D=4$. Both the $(-1)$-brane and the 
7-brane are important in string theory. First it is believed that
R-R-charged $p$-branes describe the same BPS-states as the
$Dp$-branes defined in string perturbation theory. Therefore one needs
supergravity $p$-branes for all values of $p$. Second the $(-1)$-brane
can be used to define and compute space-time instanton corrections
in string theory, whereas the 7-brane is used in the F-theory
construction of non-trivial vacua of the type IIB string.
One also expects to find R-R-charged $p$-branes with $p=8,9$.
For those values of $p$ there are no corresponding gauge fields.
The gauge field strength $F_{10}$ related to an 8-brane has been
identified with the cosmological constant in the massive version
of IIA supergravity. Finally the 9-brane is just flat space-time.

\subsection{The fundamental string}

The fundamental string solution is electrically charged
with respect to the NS-NS $B$-field. Its string frame metric
is
\be
ds^2_{Str} = H_1^{-1}(x_i) (-dt^2 + dy^2) + \sum_{i=1}^8 dx_i^2 \;,
\ee
where $H_1$ is a harmonic function with respect to the eight transverse
coordinates,
\be
\Delta^{\perp}_{x_i} H_1 = 0 \;.
\ee
Single center solutions are described by the spherically symmetric
choice
\be
H_1(r) = 1 + \frac{Q_1}{r^6} \;,
\ee
where $Q_1$ is a positive constant. 
To fully specify the solution we have
to display the $B$-field and the dilaton:
\be
B_{ty} = H_1^{-1} - 1 \mbox{   and   }
e^{-2 \phi} = H_1 \;.
\ee
All other fields are trivial. Since only NS-NS fields are excited,
this is a solution of both IIA and IIB theory. In order to
interpret the solution we have to compute its
tension and its charge. Both quantities can be extracted from
the behaviour of the solution at infinity.

The analogue of mass for a $p$-brane is the mass per world
volume, or tension $T_p$. Generalizing our discussion of 
four-dimensional asymptotically flat space-times, the tension can
be extracted by compactifying the $p$ world volume directions and
computing the mass of the resulting pointlike object in 
$d=D-p$ dimensions, using the $d$-dimensional version
of (\ref{DefMass}),
\be
g_{00} = - 1 + \frac{16 \pi G_N^{(d)}}{(d-2) \omega_{d-2}}  
\frac{M}{r^{d-3}} + \cdots \;.
\label{DefTension}
\ee
Note that this formula refers to the Einstein metric. As explained
above the standard definitions of mass and energy are tied to the
Einstein frame metric. $G_N^{(d)}$ is the $d$-dimensional Newton
constant and $\omega_{d-2}$ is the volume of the unit sphere
$S^{d-2} \subset \mathbb{R}^{d-1}$,
\be
\omega_n = \frac{ 2 \pi^{ (n+1)/2 } }{ \Gamma( \frac{n+1}{2} )} \;.
\ee
The quantity $r_S$, where
\be
r_S^{d-3} = \frac{16 \pi G_N^{(d)} M}{(d-2) \omega_{d-2}} 
\ee
has dimension length and 
is the $d$-dimensional Schwarzschild radius.

The tension of the $p$-brane in $D$ dimensions and the
mass of the compactified brane in $d$ dimensions are related by
$G_N^{(D)} T_p = G_N^{(d)} M$, because 
$G_N^{(D)} = V_p G_N^{(d)}$ and $T_p = M / V_p$,
where $V_p$ is the volume of the internal space. Dimensional
reduction of actions and branes will be discussed in some 
detail in the next section.

The $p$-brane charge is computed by measuring the flux of the
corresponding field strength through an asymptotic sphere 
in the transverse directions. Since in the limit $r \rightarrow \infty$
the $B$-field takes the form (\ref{flatelectric}), it is natural
to interpret $Q_1$ as the electric $B$-field charge. This is 
the terminology that we will adopt, but we need to make two clarifying
remarks. First note that the parameter $Q_1$ in the harmonic function
is always positive. Solutions of negative charge are desribed by
flipping the sign of $B_{ty}$, which is not fixed by the equations of 
motion. If one wants to denote the negative charge by $Q_1 <0$,
then one needs to replace in the harmonic function $Q_1$  
by $-Q_1$. Obviously it is convenient and no loss in generality
to restrict oneself to the case $Q_1>0$. The second remark is
that in the string frame NS-NS action the kinetic term of the $B$-field
is dressed with a dilaton dependend factor $e^{-2 \phi}$. 
Therefore one must include this factor in order to get a conserved
charge. But since the parameter $Q_1$ is proportional to the
conserved charge,
\be
Q_{1} \simeq \oint_7 \star ( e^{-2\phi} H ) \;,
\ee
we can take $Q_1$ to be the $B$-field charge by appropriate
choice of the normalization constant.

A similar convention can be used for general $p$-brane solutions.
As we will see later, $p$-brane solutions in $D$
dimensions are characterized by harmonic functions 
\be
H = 1 + \frac{Q_p}{r^{D-p-3}} \;.
\ee
We will always take $Q_p >0$ and refer to it as the $D$-dimensional
$p$-brane charge. With this convention the charge has
dimension $L^{D-p-3}$.

Let us next study the behaviour of the fundamental string solution
for $r \rightarrow 0$. It turns out that there is a so-called
null singularity, a
curvature singularity which coincides with an event horizont and
therefore is not a naked singularity. The fundamental string
is the extremality limit of a family of so-called black string
solutions, which satisfy the inequality
\be
T_1 \geq {\cal C}  Q_1 
\label{BogoF1}
\ee
between tension and charge, where ${\cal C}$ is a constant. Black strings
with $T_1 > {\cal C} Q_1$ have an outer event horizon and an inner horizon
which coincides with a curvature singularity. In the extreme limit
$T_1 = {\cal C} Q_1$ the two horizons and the singularity coincide and 
one obtains the fundamental string. 
As for the Reissner-Nordstrom
black hole the extreme limit is supersymmetric. Type IIA/B
supergravity has 32 real supercharges and the fundamental
string is a 1/2 BPS solution with 16 Killing spinors.
In the IIA theory, the explicit form of the Killing spinors is
\be
\epsilon_{\pm} = H_1^{-1/4} (r) \epsilon_{\pm}(\infty) \;,
\ee
with
\be
\Gamma^0 \Gamma^9 \epsilon_{\pm}(\infty) = \pm \epsilon_{\pm}(\infty)\;.
\ee
The spinors $\epsilon_{\pm}$ are ten-dimensional Majorana-Weyl 
spinors of opposite chirality.
The 9-direction is the direction along the worldvolume, $x_9=y$.

One can also find static multi-center solutions which generalize the
Majumdar-Papapetrou solution. The corresponding harmonic functions are
\be
H = 1 + \sum_{I=1}^N \frac{Q_1^I}{|\vec{x} - \vec{x}_I|^6 } \;,
\ee
where $\vec{x}, \vec{x}_I$ are eight-dimensional vectors. The positions 
$\vec{x}_I$ of the
strings in the eight-dimensional transverse space are completely 
arbitrary.

Moreover it is possible to interprete $Q_1$ within the supersymmetry
algebra. The IIA supersymmetry algebra can be extended by charges
$Z_\mu$, which transform as Lorentz vectors. 
For a fundamental string along the $y$-direction the charge (actually:
charge density) is
\be
Z_\m \simeq ( 0, \cdots 0,Q_1 ) \;.
\ee
Such charges are often called central charges, though they
are not literally central, because they do not commute with Lorentz
transformations. The extended IIA algebra takes the form
\bea
\{ Q^+_{\alpha}, Q^+_{\beta} \} &=& (P+Z)_{\m} ( C P^+ \Gamma^\m)_{\alpha
\beta} \nonumber \\
\{ Q^-_{\alpha}, Q^-_{\beta} \} &=& (P-Z)_{\m} ( C P^- \Gamma^\m)_{\alpha
\beta}  \;,\\
\nonumber
\eea
where $C$ is the charge conjugation matrix and $P^{\pm}$ projects onto
positive/negative chirality. The $p$-brane charges $Z_\m$ are not
excluded by 
the classification theorem of Haag, Lopuzanski and Sohnius, because they
are carried by field configurations which do not approach the
vacuum in all directions, but only in the transverse directions.
If one compactifies along the worldvolume directions they become
central charges in the usual sense of the lower dimensional
supersymmetry algebra.

So far we have analysed the fundamental string within the framework
of supergravity.
We now turn to its interpretation within string theory.
Although we call the fundamental string a soliton (in the broad sense
explained above) it is not a regular solution of the field equations,
but singular at $r=0$. The singularity can be interpreted 
in terms of a source concentrated at the origin. 
This source term is nothing but the  type IIA/B world sheet action itself. 

We have seen how this works in the
simplified case without gravity: The integral of the gauge field
over the world-sheet (compare (\ref{localizedsource}) to (\ref{Baction})) 
\be
\frac{1}{2 \pi \alpha'} \int_2 B = 
\frac{1}{4 \pi \alpha'}  \int_{WS} d^2 \sigma B_{\m \n}(X) \der_{\alpha} X^\m 
\der_{\beta} X^\n \ve^{\alpha \beta}
\label{Bsource}
\ee
yields upon variation a $\delta$-function source in the generalized
Maxwell equations, see (\ref{inhomeq}),(\ref{harmA}). 
Similarly the full world sheet action is the appropriate 
source for the fundamental string solution. Therefore the
fundamental string solution of the effective supergravity
theory is interpreted as describing the long range fields
outside a fundamental type IIA/B string, as already indicated
by its name. As a consequence the tension $T_1$ of the supergravity
string solution must be an integer multiple $T_1 = \wh{Q}_1 \tau_{F1}$
of the IIA/B string tension
$\tau_{F1}=\ft{1}{2 \pi \alpha'}$, where the integer $\wh{Q}_1$
counts the number
of fundamental IIA/B strings placed at $r=0$. From formula
(\ref{Bsource}) it is obvious that $\tau_{F1}$ measures
the coupling of the $B$-field to the string world sheet and therefore
it can be interpreted as the fundamental electric charge unit. This
provides a somewhat different definition of the charge than the one
by the parameter $Q_1$. Since both kinds of definition are used
in the literature, we will now explain how they are related for a 
generic $p$-brane.

Consider a $p$-brane which is charged with respect to a
$(p+1)$-form gauge potential $A_{p+1}$. The coupling
between the brane and the gauge field is described by 
\be
\tilde{Q}_p \int_{p+1} A_{p+1} \;,
\ee
where $\tilde{Q}_p$ has the dimension $L^{-p-1}$ of 
mass per worldvolume, or tension and measures the strength of 
the source. Like the parameter $Q_p$ also $\tilde{Q}_p$ measures
the conserved charge associated with the gauge field. 
But both quantities have a different dimension and therefore differ
by appropriate powers of $\alpha'$. By dimensional analysis
the relation between the (transverse) 
Schwarzschild radius $r_S$, tension $T_p$
and the charges $Q_p, \tilde{Q}_p$ is
\be
r_S^{D-p-3} \simeq Q_p \simeq G_N^{(D)} T_p \simeq G_N^{(D)} 
\tilde{Q}_p \;,
\label{relations}
\ee
up to dimensionless quantities. The $D$-dimensional Newton
constant is related to the $D$-dimensional gravitational
coupling $\kappa_D$ by $\kappa_D^2 = 8 \pi G_N^{(D)}$. 
We will see how $\kappa_D$ is related to $\kappa_{10}$
in the next section, which discusses dimensional reduction.
The relation between $\kappa_{10}$ and $\alpha'$ was given
in (\ref{kappaalpha}). The dimensionless quantities not
specified in (\ref{relations}) fall into two classes: First there
are numerical factors, which depend in part on conventional
choices like, for example, the choice of the constant in
(\ref{kappaalpha}). They are only important if the precise
numerical values of tensions of charges are relevant. We will
see two examples: the comparison of $p$-branes and $D$-branes
and the entropy of five-dimensional black holes. We will not keep
track of these factors ourselves, but quote results from the
literature when needed. The second kind of dimensionless
quantity which we surpressed in (\ref{relations})  is the 
dimensionless string coupling $g_S$. As we will see 
this dependence is very important for the qualitative behaviour
and physical interpretation of a $p$-brane.

Let us return to the specific case of the fundamental string solution,
which  carries tension and charge
$T_1 = \tilde{Q}_1 = \wh{Q}_1 \tau_{F1}$. The 
dimensionless ratio of tension and charge is independent of the string
coupling,
\be
T_1 = \tilde{Q}_1 \;.
\label{F1}
\ee
This is specific for fundamental strings. 
For a soliton (in the narrow sense)
one expects that the mass / tension is proportional to $g_S^{-2}$,
whereas (\ref{F1}) is the typical behaviour of the fundamental 
objects of a theory.

A further check of the interpretation of the fundamental string solution
is provided by looking at so-called oscillating
string solutions, which are obtained by superimposing a 
gravitational wave on the fundamental string. These solutions have
8 Killing spinors and preserve $1/4$ of the supersymmetries of
the vacuum. Similarly the perturbative IIA/B string has excitations
which sit in $1/4$ BPS representations. These are the 
states which have either only left-moving oscillations 
or only right-moving oscillations. The spectrum of such excitations 
matches precisely with the oscillating string solutions.

Finally we mention that the fundamental string solution is not
only a solution of supergravity but of the full IIA/B string theory.
There is a class of exact two-dimensional conformal field theories,
called chiral null models, which includes both the fundamental string
and the oscillating strings. This is different for the other 
supergravity $p$-branes, where usually no corresponding exact conformal field
theory is known.

\begin{Exercise}
Apply T-duality, both parallel and orthogonal to the world volume,
to the fundamental string. Use the formulae (\ref{Buscher}).
Why can one T-dualize with respect to a direction orthogonal
to the world volume, although this is not an isometry direction? 
\end{Exercise}

\subsection{The solitonic five-brane}

The solitonic five-brane (also called NS-five-brane) is magnetically
charged with respect to the NS-NS $B$-field. Again the solution is
parametrized by a harmonic function,
\be
ds^2_{Str} = -dt^2 + \sum_{m=1}^5 dy_m^2 + H_5(x_i) 
\sum_{i=1}^4 dx_i^2 \;,
\ee
\be
e^{-2 \phi} = H_5^{-1} \;,\;\;\;
H_{ijk} = \frac{1}{2}\ve_{ijkl} \der_l H_5 \;,
\ee
where
\be
\Delta_{x_i}^{\perp} H_5 = 0 \;.
\ee
For a single center solution the harmonic function is
\be
H_5 = 1 + \frac{Q_5}{r^2} \;,
\ee
where $Q_5 >0$ is the magnetic $B$-charge.
Like the fundamental string the solitonic five-brane saturates
an extremality bound. And
again there are 16 Killing spinors and static multi-center
solutions. The condition imposed on the Killing spinor of the IIA theory
is
\be
\epsilon_{\pm} = \mp \Gamma^1 \Gamma^2 \Gamma^3 \Gamma^4 \epsilon_{\pm} \;,
\ee
where $\Gamma^i$ correspond to the transverse directions $x_i$.

But this time there is no need to introduce a source term at $r=0$,
at least when probing this space-times with strings: the solution
is geodesically complete in the string frame and a
string sigma-model with this target space is well defined. 
Therefore the five-brane is interpreted as a soliton in the narrow
sense of the word, as a fully regular extended solution of the
equations of motion, like, for instance, 't Hooft-Polyakov monopoles
in Yang-Mills theories. 

The electric and magnetic $B$-field charge are subject to the
generalized Dirac quantization condition:
\be
Q_1 Q_5 \simeq  n \;.
\label{15quantization}
\ee
Therefore $Q_1,Q_5$ can only take discrete values. In the last
section we arrived at the same conclusion for $Q_1$ by a 
different reasoning, namely by identifying the source of the
fundamental string solution as the perturbative type II string.

One can introduce fundamental charge units $c_1,c_5$, which
satisfy (\ref{15quantization}) with $n=1$. Then the charges  
carried by fundamental strings and solitonic five-branes are
integer multiples of these charge units,
\be
Q_i = \wh{Q}_i c_i, \;\;\;\wh{Q}_i \in \mathbb{Z}, \;\;\;
i=1,5.
\ee
The charge unit $c_1$ is known from the identification of
the fundamental string solution with the perturbative
IIA/B string and $c_5$ is fixed by the quantization law.
The fundamental charge units are:\footnote{Here and in the
following formula we use the conventions of \cite{Mal:1996}.}
\be
c_1 = \frac{ 8 G_N^{(10)}}{6 \alpha' \omega_7} \mbox{   and   }
c_5 = \alpha'\;,
\ee
where $G_N^{(10)}$ is the ten-dimensional Newton constant, which
is related to $\alpha'$ and to the string coupling by
\be
G_N^{(10)} = 8 \pi^6 g_S^2 (\alpha')^4 \;.
\ee

One can also define a five-brane charge $\tilde{Q}_5$ which measures
the coupling of the five-brane worldvolume theory to the 
dual gauge field $\tilde{B}_6$, where $d \tilde{B}_6 = \star H$. 
When considering the $\star$-dualized
version of the NS-NS action, where $\tilde{B}_6$ is taken to
be a fundamental field instead of $B_2$, then the solution is not
geodesically complete and requires the introduction of a source term
proportional to $\int_{6} \tilde{B}_6$. As descibed in the last
section one can define a charge $\tilde{Q}_5$, which is
related to $\tilde{Q}_1$ by Dirac quantization. $\tilde{Q}_1$ and
$\tilde{Q}_5$ are
integer multiples of a charge units $\mu_{F1}$ and $\mu_{NS5}$, which measure
the electric and magnetic couplings of $B_2$ to its sources:
$\tilde{Q}_i = \wh{Q}_i \mu_{i}$, where $\wh{Q}_i \in \mathbb{Z}$.
The explicit values of the couplings 
are\footnote{These quantities correspond to $\tau_{F1}$ and
$g^2 \tau_{NS5}$ in \cite{Pol:1998}.}:
\be
\mu_{F1} = \frac{1}{2 \pi \alpha'} \mbox{   and   }
\mu_{NS5}  =  \frac{1}{ (2\pi)^5 (\alpha')^3 } \;.
\ee

The relation beween five-brane tension $T_5$, five-brane charge
$\tilde{Q}_5$ and the string coupling $g_S$ is 
\be
T_5 = \frac{\tilde{Q}_5}{g_S^2} \:,
\ee
which is the typical behaviour of the mass or tension of a 
soliton (in the narrow sense). This is consistent with
our interpretation: Fundamantal strings, which are electrically
charged under the $B$-field are the fundamental objects of the
theory. Therefore five-branes, which are magnetically charged,
should be solitons.

Finally we mention that there is an exact conformal field theory which
desribes the near horizon limit of the five-brane. It consists
of a linear dilaton theory corresponding to the transverse radius,
an $SU(2)$  Wess-Zumino-Witten model corresponding to the 
transversal three-sphere and free scalars corresponding to the
world volume directions.

\subsection{R-R-charged $p$-branes}

We now turn to the class of $p$-branes which carry R-R-charge,
restricting ourselves to the branes with $0 \leq p \leq 6$.
Again these solutions saturate an extremality bound, 
have 16 Killing spinors and admit static multi-center
configurations.

The string frame metric is
\be
ds_{Str}^2 = H_p^{-1/2}(x_i) \left( -dt^2 + \sum_{m=1}^p dy_m^2 
\right) + H_p^{1/2}(x_i) \left( \sum_{i=1}^{D-p-1} dx_i^2 \right) \;,
\ee
where $H_p$ is harmonic, $\Delta^{\perp}_{x_i} H_p = 0$.
The dilaton and R-R gauge fields are
\be
e^{-2 \phi} = H_p^{(p-3)/2} \mbox{   and   }
A_{01 \cdots p} = - \frac{1}{2}( H_p^{-1} - 1 ) \;.
\ee
When taking the $(p+1)$-form potentials with $p=0,1,2$
to be fundamental fields, then $p$-branes are electric for
$p=0,1,2$ and magnetic for $p=4,5,6$. 
For $p=3$ we have to take into account that the field strength
$F_5$ is self-dual. In this case $A_4$ 
is not a gauge potential for $F_5$. Instead $dA_4$ gives the
electric components of $F_5$ and the magnetic components are 
then fixed by self-duality.
Note that the  $D3$-brane is not only
dyonic (carrying both electric and magnetic charge)
but self-dual, i.e. electric and magnetic charge are dependent.

The electric charges $Q_p$ and the magnetic charges
$Q_{\tilde{p}}$ ($\tilde{p}=D-p-4$) are related by
the Dirac quantization condition: 
\be
Q_p Q_{\tilde{p}} \simeq n \;,\;\mbox{where}\;\;\;
n \in \mathbb{Z} \;.
\ee
Therefore the spectrum of such charges is discrete and one can
introduce fundamental charge units $c_p$:
\be
Q_p = \wh{Q}_p c_p \;\;\;\mbox{with}\;\;\;\wh{Q}_p \in \mathbb{Z} \;.
\ee
Moreover
within string theory 
R-R charged $p$-brane solutions are interrelated through 
T-duality and related to the fundamental string and to the
NS-five-brane through S-duality.\footnote{The action of T-duality
on $p$-brane solutions is discussed in some of the exercises.
S-duality transformations work in a similar way.}
Therefore the fundamental charge units are fixed and known.
The explicit values are:\footnote{Here and in the following we use
the conventions of \cite{Pol:1998}.}
\be
c_p = g_S ( \alpha')^{(7-p)/2} \frac{ (2\pi)^{6-p} }{\omega_{6-p}} \;.
\ee
The charges  $\tilde{Q}_p$, which measure the coupling
of $A_{p+1}$ to a $p$-brane source are integer multiples of
the fundamental coupling $\mu_{Dp}$, where
\be
\mu_{Dp} = \frac{1}{ (2 \pi)^p (\alpha')^{(1+p)/2} } \;.
\ee
The dependence of the tension on the charge 
$\tilde{Q}_p = \wh{Q}_p \mu_{Dp}$, where $\wh{Q}_{p} \in \mathbb{Z}$, is
\be
T_p = \frac{\tilde{Q}_p}{g_S} \;.
\ee
This behaviour is between the one of perturbative string states and
standard solitons. At this point it is important that in string theories with
both open and closed strings the closed string coupling $g_S$ and
the open string coupling $g_O$ are related as a consequence of unitarity,
\be
g_S = g_O^2 \cdot \mbox{const} \;.
\ee
Thus one should try to interprete R-R charged $p$-branes as 
solitons\footnote{Since for $p\not=3$ the solution is singular
on the horizon, these are not solitons in the narrow sense. 
As in the case of the fundamental string one needs to introduce
a source. This leads to the question whether these branes have the
same fundamental status in the theory as strings. It is possible
that M-theory is not just a theory of strings.} related
to an open string sector of type II theory. The natural way to 
have open strings in type II theory is to introduce $D$-branes.
This leads to the idea that $Dp$-branes and R-R $p$-branes describe
the same BPS states.

\subsection{$Dp$-branes and R-R charged $p$-branes}

The idea formulated at the end of the last paragraph can be
tested quantitatively by 
comparing the static R-R force at large distance between 
two $Dp$-branes to the 
one between two R-R charged $p$-branes. In both cases the full force 
is exactly zero, because these are BPS states, which admit static
multi-center configurations. One can, however, isolate the part 
corresponding to the exchange of R-R gauge bosons.

The force between two $Dp$-branes is computed in string perturbation
theory by evaluating an annulus diagram with Dirichlet boundary conditons
on both boundaries. This diagram can be viewed as a loop diagram
of open strings or as a tree level diagram involving closed strings.
In the later case the annulus is viewed as a cylinder connecting the two
$Dp$-branes. For large separation this picture is more adaequate, because
the diagram is dominated by massless closed string states. There
is an NS-NS contribution from the graviton and dilaton 
and an R-R contribution from
the corresponding gauge field $A_{p+1}$. 
The R-R contribution
gives a generalized Coulomb potential
\be
V_{R-R}(r) = \frac{Q_{Dp}^2}{r^{D-p-3}} \;,
\ee
where the charge $Q_{Dp}$ of a single $Dp$-brane is found to be
\be
Q_{Dp} = g_S ( \alpha')^{(7-p)/2} \frac{ (2\pi)^{6-p} }{\omega_{6-p}} 
= c_p \;.
\ee
Thus a single $Dp$-brane carries precisely one unit of R-R $p$-brane
charge. The long-distance R-R force between two R-R $p$-branes takes
the same form, with charge $Q_p=\wh{Q}_p c_p$. This suggests that
a R-R $p$-brane of charge $Q_p$ is build out of $\wh{Q}_p$ 
Dirichlet $p$-branes. Further tests of this idea can be performed
by studying systems of $p$-branes and $p'$-branes, with $p \not= p'$.
One can also consider the low velocity scattering of various fields
on $p$-branes and $Dp$-branes.

The $D$-brane picture and the 
supergravity 
picture have different regimes of validity. The $D$-brane picture
is valid within string perturbation theory. When considering
a system of $\wh{Q}_p$ $D$-branes, then the effective coupling is
$\wh{Q}_p g_S$, because every boundary of the worldsheet 
can sit on any of the $\wh{Q}_p$ $D$-branes. Therefore the
validity of perturbation theory requires
\be
\wh{Q}_p g_S \ll 1 \;.
\label{valpertth}
\ee
In this picture the geometry is flat and $D$-branes are 
$p$-dimensional planes, which support an open
string sector. Note that one is not restricted
in energy. Using string perturbation theory
it is possible to compute scattering processes at arbitrary energies, 
including those involving excited string states.

The range of validity of the supergravity picture is (almost)
complementary. Here we have a non-trivial curved space-time which
is an exact solution to the low energy equations of motion. In order
to reliably use the low energy effective action we are restricted 
to small curvature, measured in string units. According
to (\ref{relations}) the Schwarzschild radius, which sets the
scale of the solution is
\be 
r_S^{D-p-3}  \simeq Q_{Dp} \simeq 
\wh{Q}_p g_S (\alpha')^{(D-p-3)/2}
\label{SchwSchRR}
\ee
Therefore the condition of weak curvature,\footnote{For $p\not=3$ 
the R-R $p$-brane solutions have a null singularity. Therefore the
supergravity picture is only valid as long as one keeps a sufficient
distance from the horizon. 
The R-R 3-brane has a regular event horizon and
the same is true for certain more complicated configurations of 
intersecting branes, as we will see in the next section. For such
objects the supergravity description is valid (at least) up to 
the event horizon. The same is true if one makes the R-R 
$p$-branes with $p\not=3$ non-extremal, because in this case they
have regular horizons with a curvature controlled by $r_S$.} 
\be
r_S \gg \sqrt{\alpha'}
\ee
is equivalent to
\be
\wh{Q}_p g_S \gg 1 \;,
\ee
which is the opposite of (\ref{valpertth}). The existence 
of a dual, perturbative description of R-R charged $p$-branes
is a consequence of the particular dependence of the
Schwarzschild radius on the string coupling $g_S$. According
to (\ref{SchwSchRR}) the Schwarzschild radius goes to zero
relative to the string scale $\sqrt{\alpha'}$ when taking the
limit $g_S \rightarrow 0$. This regime is outside the 
range of the supergravity picture, but inside the regime of
perturbation theory, because
\be
r_S \ll \sqrt{\alpha'} \Leftrightarrow \wh{Q}_p g_S \ll 1 \;.
\ee
This relation also tells us why $D$-branes do not curve
space-time, although they are dynamical objects which carry
charge and mass: in the perturbative regime gravitational 
effects become arbitrarily small because the Schwarzschild
radius is much smaller than the string scale.

This is a particular feature of R-R charged $p$-branes. 
One can regard it as a consequence of the particular 
dependence of the tension on the string coupling,
\be
r_S^{D-p-3} \simeq G_N^{(D)} T_p \simeq g_S^2 \frac{1}{g_S} = g_S
\rightarrow_{g_S \rightarrow 0} 0 \;.
\ee
A completely different behaviour is exhibited
by standard solitons like the NS-five-brane. Here the gravitational
effects stay finite irrespective of the value of the coupling, because
the tension goes like $g_S^{-2}$:
\be
r_S^{D - 8} \simeq G_N^{(10)} T_{NS5} \simeq g_S^2 \frac{1}{g_S^2}
= 1 \;.
\ee
Therefore there is no perturbative description of the NS-five-brane
as a hypersurface defect in flat space-time. NS-five-branes as seen
by strings always have a non-trivial geometry.

Since the $Dp$-brane and the supergravity $p$-brane describe the same
BPS state we can interpolate between the two pictures by varying 
$g_S$ while keeping $\wh{Q}_p$ fixed. 
We have no explicit description of the intermediate regime, but we
can compute quantities in one picture and extrapolate 
to the other. This is used when counting black hole microstates 
using $D$-branes.

Before turning to this topic, we would like to digress and
shortly explain
how the Maldacena conjecture or AdS-CFT correspondence
fits into the picture.

\subsection{The AdS-CFT correspondence}

The most simple example of the AdS-CFT correspondence is provided
by a system of $N := \wh{Q}_3$ $D3$-branes. The crucial observation of
Maldacena was that there is a regime in parameter space where 
both the $D$-brane picture and the supergravity picture apply.
Let us start with the $D$-brane picture:
Since gravity is
an irrelevant interaction in $3+1$ dimensions, one can
decouple it from the world volume theory of the $D3$-brane
by taking the low energy limit
\be
\alpha' \rightarrow 0, \;\;\; \mbox{with $g_S N$ and 
$\ft{R}{\alpha'}$  fixed} \;\;,
\label{lowenergylimit}
\ee
where $R$ is the typical scale of separation of the branes.
The resulting world volume theory is four-dimensional 
$N=4$ super-Yang-Mills with gauge group $U(N)$.
$R/\alpha'$ is the typical mass scale, the W-mass.
The gauge coupling is $g_{YM}^2 = 4 \pi g_S$. After taking
the low energy limit 
one can go to large Yang-Mills coupling. This regime is of
course beyond perturbation theory, but since gravity and all
other stringy modes have been decoupled, we know that it is
still the same super Yang-Mills theory.
Besides the strong coupling limit,
$\lambda = N g_{YM}^2 \gg 1$ one can consider the
't Hooft limit $N \rightarrow \infty$ with $\lambda$ fixed.

On the supergravity side the low energy limit 
(\ref{lowenergylimit}) is a near horizon limit which maps
the whole solution onto its near horizon asymptotics.
The $D3$-brane has a smooth horizon, with geometry 
$AdS^5 \times S^5$. The supergravity picture is valid if
the curvature is small, $N g_S \gg 1$ or
$\wh{Q} g_{YM}^2 \gg 1$. In other words supergravity 
is valid in the strong coupling limit
of the gauge theory, whereas $\frac{1}{N}$ corrections
correspond to perturbative string corrections. This is
the most simple example in a series of newly proposed
'bulk - boundary'
dualities between supergravity or superstring theory 
on a nontrivial space-time and gauge theory on its
(suitably defined) boundary. Since most of the cases considered
so far relate
supergravity or string theory on AdS space to conformally
invariant gauge theories on its boundary, this is called
the AdS - CFT correspondence.
We will not enter into this subject here
and refer the interested reader to 
Zaffaroni's lectures and to the literature.

\subsection{Literature}
An extensive introduction to string theory, which also covers
$D$-branes and other more recently discovered aspects, is provided
by Polchinski's books \cite{Pol:1998}. $D$-branes were also discussed
in Gaberdiel's lectures at the TMR school. T-duality is reviewed
by Giveon, Porrati and Rabinovici
in \cite{GivPorRab:1994}. The T-duality rules for R-R fields,
which we did not write down, can be found in the paper
\cite{BerDeR:1996} by Bergshoeff and de Roo.
Our discussion of $p$-branes follows Maldacena's thesis
\cite{Mal:1996}, where more
details and references can be found.
For a detailed account on $p$-branes
in supergravity and string theory we also refer to the reviews by
Duff, Khuri and Lu
\cite{DufKhuLu:1994}, Stelle 
\cite{Ste:1997} and Townsend 
\cite{Tow:1997}. The AdS - CFT correspondence
was discussed in Zaffaroni's lectures at the TMR school.
Finally we mentioned that $D(-1)$-branes can be used to describe
instantons in string theory. This was one of the subjects in Vandoren's
lectures.

\section{Black holes from $p$-branes}
\setcounter{equation}{0}

The basic idea of the $D$-brane approach to black hole entropy is
the following: First one constructs extremal black holes by dimensional
reduction of $p$-branes. This provides an embedding of such black holes
into higher-dimensional supergravity and string theory.
Second one uses the $D$-brane description of the $p$-branes
to identify and count the states and to compute the statistical
entropy $S_{\mscr{stat}} = \log N$. The result can then be compared to the
Bekenstein-Hawking entropy $S_{BH}= \ft{A}4$ of the black hole.

This has been work out in great detail over the last years
for four- and five-dimensional extremal black holes in string 
compactifiactions with $N=8,4,2$ supersymmetry. For simplicity 
we will consider the most simple case, extremal black holes in
five-dimensional $N=8$ supergravity. This is realized by compactifying
type II string theory on $T^5$. Before we can study this example,
we have to explain how the dimensional reduction of the effective action
and of its $p$-brane solutions works.

\subsection{Dimensional reduction of the effective action}

We illustrate the dimensional reduction of actions by considering
the terms which are the most important for our purposes. The
starting point is the string frame graviton - dilaton action in $D$ dimensions,
\be
S = \frac{1}{2 \kappa^2_D} \int d^D x \sqrt{-G} e^{-2 \phi_D} \left(
R + 4 \der_M \phi_D \der^M \phi_D \right) \;.
\label{D-dim}
\ee
We take one direction to be periodic:
\be 
(x^M) = (x^\m,x)\;,\; \mbox{where} \;\;\; x \simeq x + 2 \pi R \;.
\ee
The following decomposition of the metric leads directly to
a $(D-1)$-dimensional string frame action:
\be
( G_{MN}) = \left( \begin{array}{ll}
\overline{G}_{\m \n} + e^{2 \sigma} A_\m A_\n & e^{2 \sigma} A_\m \\
e^{2 \sigma} A_\n & e^{2 \sigma} \\
\end{array} \right) \;,
\label{DimRedMetr}
\ee
where $\overline{G}_{\m\n}$ is the $(D-1)$-dimensional string frame
metric, $A_\m$ is the Kaluza-Klein gauge field and $\sigma$ is
the Kaluza-Klein scalar. Observe that the
decomposition of $G_{MN}$ is such that
\be
\sqrt{-G} = e^{\sigma} \sqrt{ - \overline{G}} \;.
\ee
Therefore the geodesic length $2 \pi \rho$ of the internal circle
and its parametric length $2 \pi R$ are related by
\be
2 \pi \rho = 2 \pi R \, e^{\la \sigma \ra} \;.
\ee
The vacuum expectation value $\la \sigma \ra$ of the Kaluza-Klein
scalar is not fixed by the equations of motion and therefore 
$\la \sigma \ra $ is a free parameter characterizing the Kaluza-Klein
vacuum. Such scalars are called moduli. Since the dilaton shows
the same behaviour it is often also called a modulus.

One should note that only the combination $\rho = R e^{\la \sigma \ra }$
has an invariant meaning, because it is the measurable, geodesic 
radius of the internal circle. One has several options of 
parametrizing the compactification. One choice is to set $R=1$ 
($R=\sqrt{\alpha'}$ when restoring units) 
and to use $\la \sigma \ra$ to parametrize 
$\rho$. The other option is to redefine $\sigma$ such that 
$\la \sigma \ra = 0$. Then the  parametric and geodesic length 
are the same, $\rho = R$. As we have seen above similar remarks
apply to the dilaton, which appears in the particular combination
$\kappa_{D} e^{\phi_D}$ with the dimensionful string coupling $\kappa_{D}$.
In the following we will use the convention that
the vacuum expectation values of the geometric moduli and of the
dilaton are absorbed in the corresponding parameters.

In order to get a $(D-1)$-dimensional string frame action it is
necessary to define the $(D-1)$-dimensional dilaton
by
\be
\overline{\phi}_{D-1} = \phi_D - \frac{\sigma}{2} \;.
\ee
One now makes a Fourier expansion of the $D$-dimensional action (\ref{D-dim})
and drops the non-constant modes which describe massive 
modes from the $(D-1)$-dimensional point of view. The resulting 
action is
\bea
S &=& \frac{1}{2 \kappa_{D-1}^2} \int d^{D-1}x \sqrt{- \overline{G}}
e^{-2 \overline{\phi}_{D-1}} \big(
\overline{R} + 4 \der_\m \overline{\phi}_{D-1} \der^\m \overline{\phi}_{D-1}
- \der_\m \sigma \der^\m \sigma   \nonumber \\
 & &  - \frac{1}{4} e^{2 \sigma}
F_{\m\n} F^{\m\n} \big) \;, \nonumber \\
\label{D-1-dim}
\eea
where the $(D-1)$-dimensional coupling is
\be
\frac{1}{\kappa_{D-1}^2} = \frac{ 2\pi R}{\kappa_D^2} \;.
\ee

Upon compactification the diffeomorphism invariance of the circle
has turned into a $U(1)$ gauge symmetry. Massive states with a non-vanishing
momentum along the circle are charged under this $U(1)$. Since the
circle is compact, the charge spectrum is discrete and is of the
form $Q \sim \frac{n}{R}$, with $n \in \mathbb{Z}$. Note that the 
gauge coupling in (\ref{D-1-dim})
is field dependent, and depends on both the dilaton and the modulus.

To reduce the full type IIA/B supergravity action one also
has to consider the tensor fields $B_{MN}$ and $A_p$ and the
fermions. We will not consider the fermionic terms here. Concerning
the various $p$-form fields we remark that the reduction of
a $p$-form on $S^1$ gives a $p$-form and a $(p-1)$-form. Often one
uses Hodge-duality to convert a $p$-form into a $(D-1-p)$-form,
in particular if $D-1-p < p$, because one wants to collect all terms
with the same Lorentz structure. For example in $D=4$ the $B_{\m\n}$
field is dualized into the universal stringy axion, whereas in
$D=5$ it is dualized into a gauge field.

\subsection{Dimensional reduction of $p$-branes}

There are two different ways of dimensionally reducing $p$-branes.
The first and more obvious way is called double dimensional 
reduction or wrapping. Here one compactifies along a world volume
direction of the brane. The reduction of a $p$-brane
in $D$ dimensions yields a $(p-1)$-brane in $D-1$ dimensions, which
is wrapped on the internal circle.

The second way is called simple dimensional reduction. This time
one compactifies a transverse direction and obtains a $p$-brane
in $D-1$ dimensions. Transverse directions are of course not
isometry directions, but here we can make use of the no-force
property of BPS branes. The idea is to first construct a periodic
array of $p$-branes along one of the transverse directions and
then to compactify.

To be specific let us split the transverse directions 
as $\vec{x} = (\underline{x},x)$. Then we form a periodic array, such
that the $n$-th $p$-brane sits at $x = 2 \pi n R$, where
$n \in \mathbb{Z}$. This array corresponds to the multi-center 
harmonic function
\be
H = 1 + \sum_{n = - \infty}^n \frac{Q}{ | \vec{x} - \vec{x}_n |^{D-p-3}} \;,
\ee
where $\vec{x}_n = ( \underline{x}, 2 \pi n R)$. Finally we impose
the periodic identification $x \simeq x + 2 \pi R$ and expand
the harmonic function in Fourier modes
\be
H = 1 + \frac{ Q}{R \; | \underline{x} |^{D-p-4}}
+ O \left(  e^{- | \underline{x} | / R} \right) \;.
\ee
Since the non-constant Fourier modes are exponentially surpressed
for small $R$, one can neglect them.

There is an alternative view of this procedure. The function
\be
H=1 + \frac{ Q }{R \;| \underline{x} |^{D-p-4}}
\ee
is a spherically symmetric harmonic function with respect
to the transverse Laplacian $\Delta^{\perp}_{\underline{x}}$ 
in $D-1$ dimensions. At the same
time it is a harmonic function of the $D$-dimensional transverse
Laplacian $\Delta^{\perp}_{\vec{x}}$
and therefore corresponds to a supersymmetric solution 
of the $D$-dimensional field equations. It has cylindrical rather
the spherical symmetry and describes, from the $D$-dimensional
point of view, a $p$-brane which has been continously smeared out along
the $x$-direction. This might be viewed as the continuum limit of
the periodic array discussed above. Such solutions are called 
delocalized $p$-branes. Since the direction in which the brane has
been smeared out has become an isometry direction, one can compactify
it.

\subsection{The Tangherlini black hole}

For simplicity we will construct five-dimensional black holes
rather then four-dimensional ones. 
It is useful to know in advance how the five-dimensional
analogue of the extreme Reissner-Nordstrom black hole looks like. This
is the Tangherlini solution, which has the Einstein frame metric
\be
ds^2_E = - H^{-2} dt^2 + H (dr^2 + r^2 d \Omega_3^2) \;.
\ee
$H$ is harmonic with respect to the four transverse 
directions. The single center function is
\be
H = 1 + \frac{Q}{r^2}  \;,
\ee
where $Q$ is the electric charge. The solution is similar to the
four-dimensional case, but with different powers of the 
harmonic function.
We are using isotropic coordinates,
and the event horizon is at $r=0$. Its area is
\be
A = 2 \pi^2 \lim_{r \rightarrow 0} ( r^3 H^{3/2} ) = 2 \pi^2 Q^{3/2} \;.
\ee
Here $2 \pi^2$ is the area of the three-dimensional unit-sphere.
With a trivial harmonic function, $Q=0$, one gets flat space and
the origin $r=0$ is just a point. But for $Q \not=0$ the metric
is non-trivial, and $r=0$ is a three-sphere.

The Bekenstein-Hawking entropy of this black hole is
\be
S_{BH}= \frac{A}{4}= \frac{\pi^2}{2} Q^{3/2} \;.
\ee

\subsection{Dimensional reduction of the $D1$-brane}

Let us now try to construct a five-dimensional black hole
by dimensional reduction of the $D1$-brane of type IIB on 
a five-torus $T^5$. To fix notation we start with the ten-dimensional
string frame metric
\be
ds^2_{Str} = H_1^{-1/2} ( dt^2 + dy^2 ) + H_1^{1/2} ( dx_1^2 + \cdots
dx_8^2)
\ee
and ten-dimensional dilaton
\be
e^{-2 \phi_{10}} = H_1^{-1} \;.
\ee
The harmonic function is
\be
H_1 = 1 + \frac{Q_1^{(10)}}{r^6} \;,
\ee
where $Q_1^{(10)}$ is the charge corresponding to the R-R two-form
in the ten-dimensional IIB action. We now compactify the
directions $x_5,\ldots,x_8,x_9=y$. The resulting string frame action
is
\be
ds^2_{Str} = - H_1^{-1/2} dt^2 + H_1^{1/2} \left( dx_1^2 + \cdots
dx_4^2 \right) \;,
\ee
where
\be
H_1 = 1 + \frac{Q_1}{r^2} \;.
\ee
Here $Q_1 := Q_1^{(5)}$ is the charge with respect to the five-dimensional
gauge field that is obtained by dimensional reduction of the ten-dimensional
R-R two-form. As we know from our previous discussion, 
the gauge kinetic terms get dressed with dilaton and moduli
dependent factors upon dimensional reduction. Thus 
$Q^{(5)} \not= Q^{(10)}_1$, and in order to know the precise
expression for $Q_1  = Q_1^{(5)}$ one needs to carefully keep
track of all the factors. We will give the explicit formula below.

The five-dimensional dilaton is 
\be
e^{-2 \phi_5} = e^{-2 \phi_{10}} \sqrt{ G_{\mscr{internal}} }
=H_1^{-1/4} \;.
\ee
To compute the Bekenstein-Hawking entropy we convert to the
Einstein frame,
\be
ds_E^2 = H_1^{-2/3} dt^2 + H_1^{1/3} ( dr^2 + r^2 d \Omega_3^2 ) \;.
\ee
The area of the event horizon is
\be
A = 2 \pi^2 \lim_{r \rightarrow 0} \left(
r^3 \sqrt{ \frac{Q_1}{r^2} } \right) = 0 \;,
\ee
and therefore the Bekenstein-Hawking entropy vanishes:
\be
S_{BH}= \frac{A}{4} = 0 \;.
\ee
The solution is degenerate: it does not have a finite event horizon.
Instead we encounter a null singularity as for the 
$p$-brane solutions discussed above.
This happens very often when constructing space-times with
event horizons in the presence of non-trivial scalars.
The scalars tend to take singular values at infinity or at
the horizon and the geometry is affected by this.
In our case the dilaton is singular at infinity and 
at the horizon. In the context of Kaluza-Klein theories such 
singularities are sometimes resolved by decompactification. This means
that the singular values of the moduli at the horizon or at infinity
indicate that the solution does not make sense
as a solution of the lower dimensional theory. In our case we indeed observe
that all internal radii either go to zero or to infinity at the horizon,
\be
R_{5,6,7,8} \rightarrow \infty \;,\;\;\;R_9 \rightarrow 0 \;.
\ee
In some cases (generically when the higher-dimensional solution 
does not have scalars) the decompactified higher-dimensional
solution is regular at the horizon. This happens for example when
considering the fundamental IIA string as a wrapped M2-brane of
eleven-dimensionsal supergravity. In our case the decompactified 
solution is the $D1$-brane, which is still singular. As discussed 
above, the singularity is interpreted in terms of a source.

We are interested in finding solutions with a regular metric 
and regular scalars. 
Both is correlated: Solutions with regular
scalars usually have regular horizons. The problem of finding
solutions with regular scalars is called the problem
of 'stabilizing the moduli'.
The generic method to achieve this is to 
construct solutions where the scalars are given by ratios
of harmonic functions. Obviously one needs more than one
harmonic functions in order to have non-constant scalars.
In terms of $D$-branes this is realized
by considering BPS superpositions
of different types of $Dp$-branes.

\begin{Exercise}
Compactify the $D1$-brane on $T^5$ and check the formulae
given in this section.
\end{Exercise}

\subsection{$Dp$-brane superpositions}

We are already familiar with the fact that BPS states admit
multi-center realizations. More generally one can also find
superpositions of different kinds of BPS states which still 
preserve part of the supersymmetry and are BPS states themselves.
In order to find and classify these states one has to find configurations
where the conditions on the Killing spinors of both 
types of BPS solutions are compatible.

We will need the special case of $Dp$-$Dp'$ superpositions
where the branes are either parallel or have rectangular intersections.
In this case one gets BPS states if the number $n$ of relative
transverse dimensions is a multiple of 4,
\be
n = 4 k \;,\;\;\;k \in \mathbb{Z} \;.
\ee
The relative transverse directions are those where one has
Neumann boundary conditions with respect to one brane and
Dirichlet boundary conditions with respect to the other.

Moreover the resulting state preserves $1/2$ of the supersymmetry
for $k=0$ and $1/4$ for $k=1$. We need to consider the second
case, which can be realized by a $D1$-brane inside a $D5$-brane,
where we wrap all world-volume directions of the $D5$-brane
on the torus.

In the ten-dimensional string frame the metric of the $D1$-$D5$
superposition is
\bea
ds^2_{Str} &=& H_1^{-1/2} H_5^{-1/2} ( - dt^2 + dy_5^2)
+ H_1^{1/2} H_5^{1/2} ( dx_1^2 + \cdots dx_4^2 ) \nonumber \\
 & & + H_1^{1/2} H_5^{-1/2} ( dy_1^2 + \cdots + dy_4^2 ) \;,\nonumber \\
\eea
with dilaton
\be
e^{-2 \phi_{10}} = \frac{H_5}{H_1} \;.
\ee
Here $t,y_5$ are the overall parallel directions, $y_1,\ldots,y_4$ the
relative transverse directions and $x_1,\ldots,x_4$ the overall transverse
directions. When taking $H_1$ ( $H_5$ ) to be trivial, we get
back the $D5$ ($D1$) solution. Thus these solutions take the form of
superpositions, despite that they solve non-linear equations
of motion. The solution has 8 Killing spinors and preserves 
$1/4$ of the supersymmetries. The conditions on the Killing spinors
are
\be
\ve_1 = \Gamma^0 \Gamma^9 \ve_2 \mbox{   and   }
\ve_1 = \Gamma^0 \Gamma^5 \cdots \Gamma^9 \ve_2 \;,
\ee
where the labeling of directions is given according to
$(x_0, x_1, \ldots, x_4, x_5 = y_1, \ldots, x_9 = y_5)$. The first 
condition is associated with the $D1$-brane, the second with the
$D5$-brane. Every condition fixes half of the supersymmetry
transformation parameters in terms of the other half and when
combining them only $1/4$ of the parameters are independent. Note that 
the ten-dimensional spinors $\ve_{1,2}$ have the same chirality,
since we are in the chiral IIB theory.

After dimensional reduction on $T^5$ the Einstein metric is found to be
\be
ds^2_E = - (H_1 H_5)^{-2/3} dt^2 + (H_1 H_5)^{1/3} ( dr^2 + r^2 
d\Omega_3^2) \;.
\ee
The area of the event horizon is
\be
A = 2 \pi^2 \lim_{r \rightarrow 0} \left( r^3 \sqrt{ \frac{Q_1 Q_5}{r^4}}
\right) = 0
\ee
and therefore the Bekenstein-Hawking entropy is still zero. 
In order to find a regular solution we have have to constuct 
a BPS superposition with yet another object.

\subsection{Superposition of $D1$-brane, $D5$-brane and pp-wave}

One way of interpreting the vanishing entropy for the $D1$-$D5$ system
is that one is looking at the ground state of this system which is
probably unique and therefore has zero entropy. Then we should look
at excited BPS states of the system. One possibility is to add momentum
along the $D1$-brane. In the supergravity picture this is realized 
by superimposing a gravitational wave. More precisely the gravitational
waves we have to consider are planar fronted gravitational waves with
parallel rays, or pp-waves for short. They are purely gravitational 
solutions of the equations of motion and do not carry charge under
the tensor fields. Instead they carry left- or right-moving momentum.
Waves with purely left- or right-moving momentum are $1/2$ 
BPS states.\footnote{pp-waves are further discussed in the 
exercises XIII and XV.}

We now consider a superpostion of $D1$-brane, $D5$-brane
and a left-moving pp-wave along the $D1$-brane. The ten-dimensional
string frame metric is
\bea
ds^2_{Str} &=& ( H_1 H_5)^{-1/2} ( -dt^2 + dy_5^2
+ (H_K -1) ( dt^2 - dy_5)^2 ) \nonumber \\
 & & + (H_1 H_5)^{1/2} ( dx_1^2 + \ldots + dx_4^2 )
+ H_1^{1/2} H_5^{-1/2} ( dy_1^2 + \ldots + dy_4^2 ) \nonumber \\
\eea
and the dilaton is
\be
e^{-2 \phi_{10}} = \frac{H_5}{H_1} \;.
\ee
The metric for a pp-wave is obtained by taking $H_1=1=H_5$. 
It depends on a harmonic function $H_K$. The presence of the
pp-wave imposes the additional conditions 
\be
\Gamma^0 \Gamma^9 \ve_1 = \ve_1 \mbox{   and   }
\Gamma^0 \Gamma^9 \ve_2 = \ve_2
\ee
on the Killing spinors. As a consequence the resulting
configuration has four Killing spinors and preserves
$1/8$ of the supersymmetries of the vacuum.

After dimensional reduction on $T^5$ we obtain the following 
Einstein frame metric:
\be
ds^2_{Str} = - ( H_1 H_5 H_K )^{-2/3} dt^2 
+ ( H_1 H_5 H_K )^{1/3} ( dr^2 + r^2 d \Omega_3^2 ) \;.
\ee
The harmonic functions are
\be
H_i = 1 + \frac{Q_i}{r^2} \;,\;\;\;i=1,5,K \;,
\ee
where $Q_1,Q_5$ are the five-dimensional charges of the
gauge fields obtained by dimensional reduction of the 
R-R two-form. The corresponding ten-dimensional charges 
are the electric and the magnetic charge of the R-R two-form.
In five dimensions electric charges are carried by zero-branes
and magnetic charges are carried by one-branes. Therefore the
five-dimensional charges $Q_1,Q_5$ are both electric. What happens
is that the R-R two-form $A_{MN}$ gives both one- and two-forms 
upon dimensional reduction. However in five dimensions the 
two-forms can be Hodge dualized into one-forms, and these
are the objects which couple locally to zero-branes. (The reduction
of the two-form on $T^5$ gives of course several one- and two-forms,
but most of them are trivial in the solution we consider.)
The parameter $Q_K$ is the related to the momentum of the 
pp-wave around the $y_5$-direction. From the lower dimensional 
point of view this is the charge with respect to one of the 
Kaluza-Klein gauge fields. All three kinds of charges are integer
multiples of unit charges, $Q_i = \wh{Q}_i c_i$, where $\wh{Q}_i \in
\mathbb{Z}$. The five-dimensional unit charges are
\be
c_1 = \frac{4 G_N^{(5)} R_9}{\pi \alpha' g_S}, \;\;\;
c_5 = g_S \alpha' \;,\;\;\;
c_K = \frac{4 G_N^{(5)}}{\pi R_9} \;.
\ee
Here $G_N^{(5)}$ is the five-dimensional Newton constant, $R_9$
the radius of the $y_5$-direction and $g_S$ is the ten-dimensional
string coupling.\footnote{To derive this one has to carefully carry
out the dimensional reduction of the action. Of course there is
a certain conventional arbitrariness in normalizing gauge fields
and charges. We use the conventions of \cite{Mal:1996}.} 
The non-trivial scalar fields of the solution are the five-dimensional
dilaton and the Kaluza-Klein scalar $\sigma$, which parametrizes the volume
of $T^5$:
\be
e^{-2 \phi_5} = \frac{ H_K^{1/2} }{ ( H_1 H_5)^{1/4} } \;,\;\;\;
e^{-2 \sigma} = \left( \frac{ H_1 }{ H_5 } \right)^{1/2} \;.
\ee
Both scalars are given by ratios of harmonic functions and are finite
throughout the solution. The Einstein frame metric is
\be
ds_E^2 = - ( H_1 H_5 H_K )^{-2/3} dt^2 + (H_1 H_5 H_K)^{1/3}
(dr^2 + r^2 d \Omega_3^2) \;.
\ee
This metric is regular at the horizon and the near horizon geometry
is $AdS^2 \times S^3$.
The area is
\be
A = 2 \pi^2 \lim_{r \rightarrow 0} \left( r^3 \sqrt{ \frac{Q_1 Q_5 Q_K}{r^6}}
\right) = 2 \pi^2 \sqrt{Q_1 Q_5 Q_K} \;.
\ee
Thus we now get a finite Bekenstein-Hawking entropy. It is instructive
to restore dimensions and to express the entropy in terms of 
the integers $\wh{Q}_i$:
\be
S_{BH} = \frac{A}{4 G_N^{(5)}} = \frac{\pi^2}{2 G_N^{(5)}}
\sqrt{ Q_1 Q_5 Q_K } = 2 \pi \sqrt{ \wh{Q}_1 \wh{Q}_5 \wh{Q}_K } \;.
\ee
All dimensionful constants and all continuous parameters cancel
precisely and the entropy is a pure number, which is given
by the numbers of $D1$-branes, $D5$-branes and quanta of 
momentum along the $D1$-branes. This indicates that an interpretation
in terms of microscopic $D$-brane states is possible. We will come
to this later.

In contrast to the entropy the mass depends on both the charges and on
the moduli. The minimum of the mass as a function of the moduli
is obtained when taking the scalar fields to be constant,
\be
e^{-2 \phi_5} = 1 \mbox{   and   } e^{-2 \sigma} = 1 \;.
\ee
This amounts to equating all the charges and all the harmonic functions:
\be
Q = Q_1 = Q_5 = Q_K , \;\;\;
H = H_i = 1 + \frac{Q}{r^2}
\ee
The resulting solution is precisely the Tangherlini solution. Solutions
with constant scalars are called double extreme. They can be deformed
into generic extreme solutions by changing the values of the moduli
at infinity. It turns out that the values of the moduli at the horizon
cannot change, but are completely fixed in terms of the charges of
the black hole. This is referred to as fixed point behaviour. 
The origin of this behaviour is that the 
stabilization of the moduli (i.e. a regular solution)
is achieved through supersymmetry enhancement
at the horizon: whereas the bulk solution has four Killings spinors,
the asymptotic solution on the event horizon has eight. 
The values of the scalars at the horizon have to satisfy 
relations, called the stabilization equations, which fix them
in terms of the charges. This has been called
the supersymmetric attractor mechanism, because the values of the
scalars are arbitrary at infinity but are attracted to their
fixed point values when going to the horizon. The geometry
of the near horizon solution is $AdS^2 \times S^3$.

\begin{Exercise}
Compactify the $D$-dimensional pp-wave, $D>4$,
\be
ds^2_E = (K-1) dt^2 + (K+1) dy^2 - 2 K dy dt + d\vec{x}^2
\ee
over $y$. Take the harmonic function to be $K=\ft{Q}{r^{D-4}}$.
Why did we delocalize the solution along $y$? What happens 
for $D=4$? What is the interpretation of the parameter $Q$
from the $(D-1)$-dimensional point of view?
\end{Exercise}

\subsection{Black hole entropy from state counting}

We now have to analyse the black hole solution in the $D$-brane
picture in order to identify and count its microstates and to
compute the statistical entropy. The $D$-brane configuration consists
of $\wh{Q}_5$ $D5$-branes and $\wh{Q}_1$ $D1$-branes wrapped on $T^5$.
Moreover $\wh{Q}_K$ quanta of light-like, left-moving 
(for definiteness) momentum
have been put on the $D1$-branes. This is an excited BPS state and
the statistical entropy counts in how many different ways one can
distribute the total momentum between the excitations of the system.
Since the momentum is light-like, we have to look for the massless
excitations. We can perform the counting in the corner of the parameter
space which is most convenient for us, because we are considering
a BPS state.

In particular we can split the $T^5$ as $T^4 \times S^1$ and make
the circle much larger than the $T^4$. After dimensional reduction
on $T^4$ the $D$-brane system is $1+1$ dimensional, with 
compact space. At low energies the effective world volume 
theory of $\wh{Q}_p$
$Dp$-branes is a dimensionally reduced 
$U(\wh{Q}_p)$ super Yang-Mills theory.
In our case we get a two-dimensional
Yang-Mills theory with $N=(4,4)$ supersymmetry and gauge group
$U(\wh{Q}_1) \times U(\wh{Q}_5)$. The corresponding excitations
are the light modes of open strings which begin and end on
$D1$-branes or begin and end on $D5$-branes. In addition there are
open strings which connect $D1$-branes to $D5$-branes or vice versa.
The light modes of these strings provide additional hypermultiplets
in the representations $\wh{Q}_1 \times \overline{\wh{Q}}_5$ and
$\wh{Q}_5 \times \overline{\wh{Q}}_1$, where $\wh{Q}_p$ and
$\overline{\wh{Q}}_p$ are the fundamental representation and its
complex conjugate.

In order to identify the massless excitations of  this theory,
one has to find the flat directions of its scalar potential. 
The potential has a complicated valley structure.
There are two main branches, called the Coulomb branch and the
Higgs branch. The Coulomb branch is parametrized by vacuum 
expectation values of scalars in vector multiplets, whereas
the Higgs branch is parametrized by vacuum expectation values of
scalars in hypermultiplets. Note that once a massless excitation along
the Coulomb branch (Higgs branch) has been turned on, all excitations
corresponding to fundamental (adjoint) scalars become massive and their
vacuum expectation values have to vanish. The two kinds of massless
excitations mutually exclude one another. Since we expect that the
state with maximal entropy is realized, we have to find out which of
the two branches has the higher dimension.

Along the Coulomb branch the gauge group is broken to the
$U(1)^{\wh{Q}_1} \times U(1)^{\wh{Q}_5}$. The number
of massless states is proportional to the number 
$\wh{Q}_1 + \wh{Q}_5$ of directions along the Cartan subalgebra.
Geometrically, turning on vacuum expectation values of the adjoint
scalars corresponds to moving all the $D1$-branes and $D5$-branes
to different positions. Then only open strings which start and end 
on the same brane can have massless excitations. The state describing
the black hole is expected to be a bound state, where all the 
branes sit on top of each other. This is not what we find in the
Coulomb branch.

Along the Higgs branch the gauge group is broken to $U(1)$. 
Since the vacuum expectation values of all adjoint scalars, which encode
the positions of the branes, are frozen to zero, all branes sit
on top of each other and form a bound state, as expected for a black hole.
The unbroken $U(1)$ corresponds to the overall translational degree 
of freedom of the bound state. A careful analysis shows that the
potential has $4 \wh{Q}_1 \wh{Q}_5$ flat directions, corresponding
to scalars in $\wh{Q}_1 \wh{Q}_5$ hypermultiplets. The massless degrees
of freedom are the $4 \wh{Q}_1 \wh{Q}_5$ scalars and the 
$2 \wh{Q}_1 \wh{Q}_5$ Weyl spinors sitting in these multiplets.

If we take the circle to be very large, then the energy carried 
by individual excitations is very small. Therefore we only need to
know the IR limit of the effective theory of the massless modes.
The $N=(4,4)$ supersymmetry present in the system implies that
the IR fixed point is a superconformal sigma-model with a 
hyper K\"ahler target space. Therefore the central charge can 
be computed as if the scalars and fermions were free fields.
Since a real boson (a Majorana-Weyl fermion)
carries central charge $c=1$ ($c=\frac{1}{2}$), the total central
charge is
\be
c = \left( 1 + \frac{1}{2} \right) 4 \wh{Q}_1 \wh{Q}_5 =
6 \wh{Q}_1 \wh{Q}_5 \;.
\ee

We now use Cardy's formula for the asymptotic number $N(E)$ of
states in a two-dimensional conformal field theory with compact
space:
\be
N(E) = \exp S_{\mscr{Stat}} \simeq \exp \sqrt{
\pi c E L /3} \;,
\ee
where $E$ is the total energy and $L$ the volume of space.
The formula is valid asymptotically for large $E$.
Using that 
\be
E = \frac{| \wh{Q}_K |}{R} \mbox{   and   }
L = 2 \pi R \;,
\ee
we find
\be
S_{\mscr{Stat}} = 2 \pi \sqrt{ \frac{c}{6} \wh{Q}_K } =
2 \pi \sqrt{ \wh{Q}_1 \wh{Q}_5 \wh{Q}_K } \;,
\ee
which is precisely the Bekenstein-Hawking entropy of the black hole.

\subsection{Literature}

Our treatment of dimensional reduction follows 
Maldacena's thesis \cite{Mal:1996} and the books
of Polchinski \cite{Pol:1998} and Behrndt \cite{Beh:2000}.
The derivation of the statistical entropy through counting
of $D$-brane states is due to Strominger and Vafa \cite{StrVaf:1996}.
The exposition given in the last section follows \cite{Mal:1996}.
The $D1-D5$ system has been the subject of intensive study since then,
see for example \cite{Dav:1999} or \cite{Man:2000} for recent reviews 
and references.
The supersymmetric attractor mechanism was discovered by
Ferrara, Kallosh and Strominger \cite{FerKalStr:1995}. 

\subsection{Concluding Remarks}

With the end of these introductory lectures we have reached
the starting point of the recent research work on black holes in
the context of string theory. Let us briefly indicate some
further results and give some more references. 

The above example of a matching between the
Bekenstein Hawking entropy and the statistical entropy was for
a five-dimensional black hole in $N=8$ supergravity. This has
been generalized to compactifications with less supersymmetry
and to four-dimensional black holes. The most general set-up where
extremal black holes can be BPS solitons are four-dimensional
$N=2$ compactifications. The Bekenstein-Hawking entropy for
such black holes was found by Behrndt et al \cite{BehCardWKalLueMoh:1996},
whereas the corresponding state counting was performed
by Maldacena, Strominger and Witten \cite{MalStrWit:1997}
and by Vafa \cite{Vaf:1997}. 
To find agreement between the geometric entropy computed in
four-dimensional $N=2$ supergravity and the statistical
entropy found by state counting one must properly include
higher curvature terms on the supergravity side and 
replace the Bekenstein-Hawking area law by a refined
definition of entropy, which is due to Wald and 
applies to gravity actions with higher curvature terms 
\cite{CardWMoh:1998/12}. An upcoming paper of the 
author will provide a detailed 
review of black hole entropy in $N=2$ compactifications,
including the effects of higher curvature terms \cite{MohHabil}.

One aspect of black hole entropy is the dependence of 
the entropy on the charges. In the example discussed above
the five-dimensional black hole carried three charges,
whereas the the most general extremal black hole solution
in a five-dimensional $N=8$ compactification carries  
27 charges. It is of considerable interest to find the most
general solution and the corresponding entropy, not only
as a matter of principle but also because string theory
predicts an invariance of the entropy formula under discrete
duality transformations. Depending on the compactification
these are called U-duality, S-duality or T-duality. In $N=8$
and $N=4$ compactifications these symmetries are exact and can be used to
construct general BPS black hole solutions from a generating
solution. Duality
properties of entropy formulae and of solutions are reviewed by D'Auria
and Fr\'e \cite{DauFre:1998}. 
More recently, a generating solution
for regular BPS black holes in four-dimensional 
$N=8$ compactifications has
been constructed by Bertolini and Trigiante, which allows
for both a macroscopic and microscopic interpretation
\cite{BerTri}. We refer to these works for more references
on generating solutions.

In heterotic $N=2$ compactifications T-duality is preserved
in perturbation theory. Since the low energy effective action receives
both loop and $\alpha'$ corrections, the situation is more complicated
than in $N=4,8$ compactifications. Nevertheless one can show that
the entropy is $T$-duality invariant and one can find explicit
$T$-duality invariant entropy formulae in suitable limits in 
moduli space.
This is discussed in \cite{CardWMoh:1999/06} and will be reviewed in 
\cite{MohHabil}.

One can also construct explicit general multi-center 
BPS black hole solutions
in four- and five-dimensional $N=2$ supergravity, which
are parametrized by harmonic functions and generalize the
Majumdar-Papapetrou solutions and the stationary IWP solutions.
These were found, for the four-dimensional case, by
Sabra \cite{Sab:1997/03}
and by Behrndt, L\"ust and Sabra \cite{BehLueSab:1997}.

Another direction is to go away from the BPS limit and to 
study near-BPS states. The most prominent application is the
derivation of Hawking radiation from the $D$-brane perspective.
Near-BPS states can for example be desribed by adding a small admixture
of right-moving momentum to a purely left-moving BPS state.
In the $D$-brane picture left- and right-moving open strings 
can interact, form closed string states and leave the brane. 
The resulting spectrum quantitatively agrees, when averaging over
initial and summing over final states, with thermal Hawking radiation.
Hawking radiation and other aspects of near-extremal black holes are
reviewed in \cite{Mal:1996}, \cite{Dav:1999}, \cite{Man:2000}
and \cite{AhaEtAl:1999}.

A second way to go away from the BPS limit is to deform multi-center
BPS solutions by giving the black holes a small velocity.
To leading order, such systems are completely determined by
the metric on the moduli space of multi-center solutions.
The quantum dynamics of such a system, including interactions
of black holes can be studied in terms
of  quantum mechanics on the moduli space. In the so-called near
horizon limit the quantum dynamics becomes superconformal and it seems
possible to learn about black hole entropy in terms of bound
states of the conformal Hamiltonian. This subject is reviewed
in \cite{BriMicStrVol:1999}.

Finally we would like to point out that there are other 
approaches to black holes in string theory. An older idea
is the identification of black holes with excited elementary
string states. This can be formulated in terms of a correspondence
principle and is reviewed in \cite{Pol:1998}. Schwarzschild
black holes have been discussed using the Matrix formulation of
M-theory \cite{BanFisKleSus:1997}. Another approach is to map
four- and five-dimensional black holes to three-dimensional black holes
through T-duality transformations which are asymptotically light-like.
One then uses that three-dimensional gravity has no local degree
of freedoms, and counts microstates in a two-dimensional conformal
field theory living on the boundary of space-time. This is 
for example reviewed in \cite{Ske:1999}. Finally the developement of the
AdS-CFT correspondence is closely interrelated with various aspects of
black hole physics \cite{AhaEtAl:1999}.
All these approaches are not tied to BPS states. This opens the
perspective that a satisfactory quantitative understanding 
of non-supersymmetric black holes will be achieved in the future.

\subsection*{Acknowledgements}

I would like to thank the organisers of the TMR 2000 school
for providing a very pleasent environment during the school.
My special thanks goes to all participants who gave feedback
to the lectures during the discussion and exercise sessions.

These lecture notes are in part also based on lectures given
at other places, including the universities of Halle, Hannover, Jena 
and Leipzig and the universiti\'e de la mediterrane, Marseille. 

Finally I would like to thank the Erwin Schr\"odinger International Institute 
for Mathematical Physics 
for its hospitality during the final stages of this work.

\begin{appendix}

\section{Solutions of the exercises}
\setcounter{equation}{0}

\begin{Solution}
For a radial lightray we have $ds^2=0$ and $d\theta=0=d\phi$.
Thus:
\be
ds^2 = - \left( 1 - \frac{r_S}{r} \right) dt^2
+ \left( 1 - \frac{r_S}{r} \right)^{-1} dr^2 = 0 
\ee
or
\be
\left( \frac{dt}{dr} \right)^2 = \left( 1 - \frac{r_S}{r} \right)^{-2}\;.
\ee
We take the square root:
\be
\frac{dt}{dr} = \pm \left( 1 - \frac{r_S}{r} \right)^{-1} \;.
\ee
The $+$ sign corresponds to outgoing lightrays, the $-$ sign
to ingoing ones. By integration we find the time intervall
\be
\Delta t = t_2 - t_1 = \pm \left[ (r_2 - r_1) + r_S
\log  \frac{ r_2 - r_S }{ r_1 - r_S } \right] \;.
\ee
In the limit $r_1 \rightarrow r_S$ the time intervall diverges,
$\Delta t  \rightarrow \infty$.
\end{Solution}

\begin{Solution}
The frequency $\omega^\m_i$ is given by the time component
of the four-momentum $k^\m$ with respect to the frame 
of a static observer at $r_i$. The time-direction of this frame is
defined by the four-velocity $u^\m_i$, which is the normalized
tangent to the world line, $u^\m_i u_{\m i} = -1$. Therefore the
frequency is
\be
\omega_i = - k_\m u^\m_i \;.
\ee
The sign is chosen such that $\omega_i$ is positive when
$k^\m, u^\m_i$ are in the forward light cone.

Next we prove equation (\ref{EnergyConservation}). $t^\m$ is 
the tangent of a geodesic and $\xi^\m$ is a Killing vector field.
The product rule implies:
\be
t^\m \nabla_\m ( \xi_\n t^\n ) =
t^\m (\nabla_\m \xi_\n)  t^\n + t^\m \xi_\n \nabla_\m t^\n \;.
\ee
The first term vanishes by the Killing equation,
$\nabla_{(\m} \xi_{\nu)} =0$, whereas the second term is zero
as a consequence of the geodesic equation $t^\m \nabla_\m t^\n =0$.

In the context of our exercise the Killing vector field is the
static Killing vector field $\xi = \ft{\der}{\der t}$ and 
the tangent vector is the four-momentum of the light ray,
$k^\m$. The conservation law implies
\be
(\xi_\n k^\n)_{r_1} = (\xi_\n k^\n)_{r_2} \;.
\ee

The conserved quantity is the projection of the
four-momentum on the direction defined by the timelike 
Killing vector field. 
Therefore it is interpreted as energy.
More generally energy is conserved along geodesics if a metric
has a timelike Killing vector.

Finally the vectors $u^\m_i$ and $\xi^\m$ are parallel: The 
observers at $r_i$ are static which means that their time-directions
coincide with the Schwarzschild time. The constant of proportionality
is fixed by the normalization: By definition four-velocities
are normalized as $u^\m_i u_{\m i} = -1$ whereas the norm of the
Killing vector $\xi^\m = (1,0,0,0)$ is
\be
\xi^\m \xi_\m = g_{\m \n} \xi^\m \xi^\n =
g_{tt} = - \left( 1 - \frac{r_S}{r} \right) \;.
\ee
Therefore
\be
\xi^\m = V(r_i) u^\m_i \;,
\ee
where $V(r_i) = \sqrt{-g_{tt}}$ is the redshift factor.
 
Putting everything together we find
\be
\frac{\omega_1}{\omega_2} = \frac{k_\m u^\m_1}{k_\m u^\m_2}
= \frac{V(r_2)}{V(r_1)} \frac{ k_\m \xi^\m }{ k_\m \xi^\m }
= \frac{V(r_2)}{V(r_1)} \;.
\ee
\end{Solution}

\begin{Solution}
We use the relation between the four-acceleration $a^\m$, the
four-velocity $u^\m$, the Killing vector $\xi^\m$ and
the redshift factor:
\be
a^\m = \frac{d u^\m}{d  \tau} = u^\n \nabla_\n u^\m
= \frac{\xi^\n}{V} \nabla_\n \frac{ \xi^\m }{V} 
= \frac{\xi^\n}{ \sqrt{-\xi_\rho \xi^\rho} } 
\nabla_\n \frac{ \xi^\m }{ \sqrt{ -\xi_\rho \xi^\rho } } 
\;.
\ee
The formulae (\ref{amu}) and (\ref{a}) follow by working
out the derivatives and making use of the Killing equation
$\nabla_{(\m} \xi_{\n)}=0$.

To derive (\ref{SurGraSch}) we compute
\be
\der_r V = \der_r \sqrt{ - \xi_\n \xi^\n } =
\frac{r_S}{2 r^2 \sqrt{ 1 - \frac{r_S}{r}} }\;,
\ee
\be
\nabla_\m V \nabla^\m V = g^{rr} ( \der_r V)^2 =
\frac{r_S^2}{4 r^4}
\ee
and finally find
\be
\kappa_S = ( Va )_{r=r_S} = \left. \sqrt{ \nabla_\m V \nabla^\m V}
\right|_{r=r_S} = \left. \frac{r_S}{2r^2} \right|_{r=r_S} =
\frac{1}{2 r_S} = \frac{1}{4M} \;.
\ee
\end{Solution}

\begin{Solution}

If $F_{\m\n}$ is static and spherically symmetric, then the independent
non-vanishing components are $F_{tr}(r,\theta,\phi)$ and
$F_{\theta \phi}(r,\theta,\phi)$. Next we note that
\be
\nabla_\m F^{\m \nu}  = \frac{1}{\sqrt{-g}} \der_{\m} \left( 
\sqrt{-g} F^{\m \n} \right) \;.
\ee
This identity is generally valid for the covariant divergence of
antisymmetric tensors of arbitrary rank. For a metric of the form
(\ref{MetricStaticSpheric}) we have
\be
\sqrt{-g} = e^{g+f} r^2 \sin \theta \;.
\ee
The only non-trivial equation of motion for the electric part is
\be
\der_r \left( \sqrt{-g} F^{rt} \right) =
\der_r \left( e^{g+f} r^2 \sin \theta \cdot (- e^{-2g -2f} F_{rt} \right)
= 
\der_r \left( e^{-g-f} r^2 \sin \theta F_{tr} \right) =0 \;,
\ee 
which implies 
\be
F_{tr} = e^{g+f} \frac{q(\theta,\phi)}{r^2} \;.
\ee
But we also have to impose the Bianchi identities
$\ve^{\m\n\rho \sigma} \der_\n F_{\rho \sigma}=0$, which imply
$\der_{\theta} F_{tr} = 0 = \der_{\phi} F_{tr}$ and therefore
we have
\be 
F_{tr} = e^{g+f} \frac{q}{r^2} 
\ee 
with a constant $q$.

The non-trivial equations of motion for the magnetic part are
\bea
\der_{\theta} \left( e^{g+f} r^2 \sin \theta F^{\theta \phi}\right) &=&0 \;,
\nonumber \\
\der_{\phi} \left( e^{g+f} r^2 \sin \theta F^{\phi \theta} \right) &=&0  \;,\\
\nonumber
\eea
which are solved by
\be
F_{\theta \phi} = p(r) \sin \theta \;.
\ee
Since the Bianchi identities imply $\der_r F_{\theta \phi} = 0$
we finally have
\be
F_{\theta \phi} = p \sin \theta \;,
\ee
with constant $p$.
\end{Solution}

\begin{Solution}
Choose the integration surface to be a sphere $r=const$ with 
sufficiently large $r$.
Introduce coordinates $y^{\alpha} = (\theta,\phi)$ on this sphere.
Then
\be
\oint {}^{\star} F = \oint \frac{1}{2} {}^{\star} F_{\m \n} 
dx^\m \wedge dx^\n = \oint \frac{1}{2} {}^{\star}F_{\alpha \beta}
dy^{\alpha} \wedge dy^{\beta} = \int_0^{\pi} d \theta \int_0^{2 \pi} d\phi
\;\; {}^{\star} F_{\theta \phi} \;.
\ee
Evaluate the $\star$-dual:
\be
{}^{\star} F_{\theta \phi} = \sqrt{-g} F^{tr} \;.
\ee
(The Levi-Civita tensor $\ve_{\m \n \rho \sigma}$
in the definition (\ref{dualgaugefield}) contains a factor
$\sqrt{-g}$.) Finally plug in $F^{tr}$ and $\sqrt{-g}$:
\be
\oint {}^{\star} F = \int_0^{\pi} d \theta \int_0^{2\pi} d\phi
e^{-f-g} \frac{q}{r^2} e^{f+g} r^2 \sin \theta  = 4 \pi q\;.
\ee
The second integral is
\be
\oint F = \int_0^{\pi} d \theta \int_0^{2\pi} F_{\theta \phi}
=\int_0^{\pi} d \theta \int_0^{2\pi} p \sin \theta
= 4 \pi p \;.
\ee
\end{Solution}

\begin{Solution}
The fields $F_{tr}$ and $F_{\theta \phi}$ look different, because
we use a coordinate system where the tangent vectors to the
coordinate lines are not normalized. To get a more symmetric
form we can use the vielbein to convert the curved indices
$\m,\n=t,r,\theta,\phi$ into flat indices $a,b=0,1,2,3$.
The natural choice of the vielbein for a spherically
symmetric metric (\ref{MetricStaticSpheric}) is
\be
e_\m^{\;\;a} = \mbox{diag} ( e^g, e^f, r, r \sin \theta ) \;,\;\;\;
e_a^{\;\;\m} = \mbox{diag} (e^{-g} , e^{-f}, \ft1r, 
\ft1r \sin^{-1} \theta ) \;.
\ee
Now we compute $F_{ab} = e_a^{\;\;\m} e_b^{\;\;\n} F_{\m \n}$
we the result
\be
F^{01} = \frac{q}{r^2} \;,\;\;\;\
F^{23} = \frac{p}{r^2}  \;.
\ee
Thus when expressed using flat indices the gauge field looks
like a static electric and magnetic point charge in flat space.
\end{Solution}

\begin{Solution}
The coordinate transformation is $r \rightarrow r - M$. 
In the limit $r\rightarrow 0$ of (\ref{extremeisotropic}) the
$(\theta,\phi)$ part of the metric is
\be
\left( 1 + \frac{M}{r} \right)^2 r^2 d\Omega^2 
\rightarrow_{r\rightarrow 0} \frac{M^2}{r^2} r^2 d \Omega^2 =
M d \Omega^2 \;.
\ee
Thus integration over $\theta, \phi$ at $r=0$ and arbitrary $t$
yields $A=4 \pi M^2$, which is the area of a sphere of radius $M$.

To show that (\ref{nearhorizon}) is conformally flat, we introduce
a new coordinate $\rho$ by $\rho/M = M/r$. Then the metric takes
the form
\be
ds^2 = \frac{M^2}{r^2} \left( - dt^2 + d \rho^2 + \rho^2 d \Omega^2 \right)\;.
\ee
This is manifestly conformally flat, because the expression in brackets
is the flat metric, written in spheric coordinates.
\end{Solution}

\begin{Solution}
The non-trivial equations of motion for $F_{ti}=\mp \der_i e^{-f}=F^{it}$ 
are
\be
\der_i \left( \sqrt{-g} F^{it} \right) = 0  \;,
\ee
with $\sqrt{ -g} = e^{2f}$, where $f=f(\vec{x})$. 
Plugging in the ansatz we get
\be
\sum_i \der_i \left( e^{2f} \der_i e^{-f} \right) =
\sum_i \der_i \der_i e^f =0 \;.
\ee
Thus $e^f$ must be a harmonic function.
\end{Solution}

\begin{Solution}
In order to one single equation for the background, the two terms
must be linearly dependent. Up to a phase this leads to the ansatz
\be
\epsilon_A = - \g^0 \ve_{AB} \epsilon^B \;.
\ee
We will comment on the significance of the phase later. Plugging this
ansatz into the Killing equation we get
\be
F^-_{0i} = \frac{1}{2} \der_i e^{-f} \;.
\ee
Since the right hand side is real it follows
\be
F_{0i} = \der_i e^{-f} \;\;\; F_{ij} =0 \;.
\ee
This solution is purely electric. If we take a different phase 
in the ansatz, we get a dyonic solution instead. Finally the Maxwell
equations imply that $e^f$ must be harmonic,
\be
\Delta e^f = 0 \;.
\ee
This is precisely the Majumdar-Papapetrou solution.
\end{Solution}

\begin{Solution}
When decomposing the Majorana supercharges in term of Weyl spinors
\be
Q^A_a = \left( \begin{array}{c} 
Q^A_{\alpha} \\ \overline{Q}^{A \dot{\alpha}} \\
\end{array} \right) \;,
\ee
one finds that
\be
\left( \begin{array}{cc}
\{ Q^1_{\alpha}, Q^2_{\beta} \} & \{ Q^1_{\alpha}, \ov{Q}^2_{\dot{\beta}} 
\}\\
\{ \ov{Q}^{2 \dot{\alpha} }, Q^2_{\beta} \} & \{ \ov{Q}^{2 \dot{\alpha}}, 
\ov{Q}^2_{\dot{\beta}}  \}\\
\end{array} \right) \sim \;\;\g^\m P_\m + i p \, \mathbb{I} + q \g_5 \;.
\ee
\end{Solution}

\begin{Solution}
We have to solve
\be
( \g^\m P_\m + ip + q \g_5 ) \epsilon =0\;.
\ee
For a massive state at rest we have
\be 
\g^\m P_\m = - M \g^0 \;.
\ee
Decomposing the Dirac spinor $\epsilon$ into chiral spinors,
\be
\epsilon = \epsilon_+ + \epsilon_- \mbox{  where   }
\g_5 \epsilon_{\pm} = \pm \epsilon_{\pm} \;,
\ee
we get
\be
M \epsilon_- + i \g^0 (p-iq) \epsilon_+ = 0 \;.
\ee
Now we use that $Z = p-iq$ is the central charge. Moreover, for a
BPS state we have $Z = M \ft{Z}{|Z|}$. This yields
\be
\epsilon_- + i \gamma^0 \frac{Z}{|Z|} \epsilon_+ = 0 \;,
\ee
which is the same projection that one finds when solving the
Killing spinor equations for the extreme Reissner-Nordstrom black hole.
\end{Solution}

\begin{Solution}
The covariante derivative is defined by
\be
D_{\pm} X^1 = \der_{\pm} X^1 + A_{\pm} \;,
\ee
where the gauge field $A_{\pm}$ transforms as
\be
A_{\pm} \rightarrow A_{\pm} - \der_{\pm} a
\ee
under $X^1 \rightarrow X^1 + a(\sigma)$. The non-vanishing
component of the corresponding field strength is 
$F_{+-} = \der_+ A_- - \der_- A_+$. In the locally invariant
action (\ref{local}) we have imposed that this field strength
is trivial using a Lagrange multiplier $\tilde{X}^1$. Using
the equation of motion $F_{+-}=0$ and choosing the gauge
$A_{\pm}=0$ we get back the original action (\ref{global}).
(We are ignoring global aspects.) Eliminating the gauge field
through its equation of motion results in
\be
S[\tilde{G}_{11}] = \int d^2 \sigma \tilde{G}_{11}
\der_+ \tilde{X}^1 \der_- 
\tilde{X}^1 \;,
\ee
where $\tilde{G}_{11} = \ft1{G_{11}}$.

Thus, by the field redefinition $X^1 \rightarrow \tilde{X}^1$ we
have inverted the target space metric along the isometry
direction. If the 1-direction is compact this means that 
string theory on a circle with radius $R$ is equivalent to
string theory on a circle with the inverse radius $R^{-1}$.
In the non-compact case small and large curvature are related.
This is the simplest example of T-duality. In the example
we have derived it from the world-sheet perspective, and 
for curved target spaces with isometries. This formulation
is also known as Buscher duality.

\end{Solution}

\begin{Solution}
We first apply T-duality along the world-volume direction $y$. 
Using (\ref{Buscher}) we get
\be
d{s'}^2_{Str} = (K-1) dt^2 + (K+1) dy^2 - 2 K dt dy + d\vec{x}^2 \;,
\label{ppwave}
\ee
where $K = H_1 -1$ and 
\be
\phi' = 0 \Rightarrow d{s'}^2_{Str} = d{s'}^2_E \;.
\ee
The non-trivial $B$-field of the fundamental string has been transformed
into an off-diagonal component of the metric. Moreover the dilaton
is trivial in the new background, which therefore is purely gravitational.
(\ref{ppwave}) is a special case of a gravitational wave (pp-wave).
Introducing light-cone coordinates $u=y-t$ and $v=y+t$ one gets
the standard form
\be
d{s'}^2_E = du dv + K du^2 + d\vec{x}^2 \;.
\ee
When T-dualizing a single center fundamental string, the 
function $K$ takes the form $K= \frac{Q_1}{r^6}$. More generally,
pp-waves are $1/2$ BPS states if $K$ has an arbitrary dependence
on $u$ and is harmonic with respect to the transverse coordinates,
\be
\Delta_{\vec{x}} K(u,\vec{x}) = 0 \;.
\ee
The solution has a lightlike Killing vector $\der_v$. The 
supersymmetry charge carried by it is its left-moving (or right-moving)
momentum. The non-trivial $u$-dependence can be used to modulate
the amplitude of the wave. For the T-dual of the fundamental string
the amplitude is constant and the total momentum is infinite unless
one compactifies the $y$-direction.

When superimposing a pp-wave with non-trivial $u$-dependence
on a fundamental string solution, one gets various oscillating string
solutions which are $1/4$ BPS solutions. These solutions are in
one to one correspondence with $1/4$ BPS states of the perturbative
type IIA/B string.

Let us next apply T-duality orthogonal to the world volume. 
Since these directions are not isometry directions, a modification
of the procedure is needed. One first delocalizes the string along
one of the directions, say $x_8$. This is done by dropping
the dependence of the solution on the corresponding coordinate:
\be
H_1(r) = 1 + \frac{Q_1}{r^6} \rightarrow H_1(r') = 1 + \frac{Q_1}{ {r'}^5}\;,
\ee
where
\be
r = \sqrt{x_1^1 + \cdots + x_8^2} \;\;\;\mbox{and}\;\;\;
r' = \sqrt{x_1^1 + \cdots + x_7^2} \;.
\ee
Note that when replacing $H_1(r)$ by $H_1(r')$ we still have a solution,
because $H_1(r')$ is still harmonic. However we have gained one
translational isometry (and lost some of the spherical symmetry in
exchange). We can now use (\ref{Buscher}) to T-dualize over
$x_8$. The resulting solution takes again the form of a fundamental
string solution and we can 'localize' it by making the inverse
replacement $H_1(r') \rightarrow H_1(r)$. Since T-duality
relates type IIA to type IIB we have mapped the fundamental
IIA/B string to the IIB/A string. 

T-duality parallel to the worldvolume relates the fundamental
IIA/B string to the IIB/A pp-wave, as discussed above. One can
also show that T-duality orthogonal to the world volume maps
the IIA pp-wave to the IIB pp-wave and vice versa. Finally 
parallel T-duality maps a $Dp$-brane to a $D(p-1)$-brane
and orthogonal T-duality maps a $Dp$-brane to $D(p+1)$-brane.
In most of these cases T-duality has to be combined with
localization and/or delocalization. In the case of $Dp$-branes
one has to know the transformation properties of R-R gauge fields
under T-duality, see \cite{BerDeR:1996}.

\end{Solution}

\begin{Solution}
The answers are given in the text.
\end{Solution}

\begin{Solution}
Using the formula (\ref{DimRedMetr}) 
for dimensional reduction we get the metric
\be
d\overline{s}^2 = - H^{-1} dt^2 + d \vec{x}^2 \;,
\ee
the Kaluza-Klein gauge field
\be
A_t = H^{-1} -1 
\ee
and the Kaluza-Klein scalar
\be
e^{2 \sigma} = H \;,
\ee
where
\be
H = K + 1 = 1 + \frac{Q}{r^{D-4}} \;.
\ee
In $D$ dimensions, $Q$ is the lightlike momentum along the
$y$-direction, whereas from the $(D-1)$-dimensional point of view
$Q$ is the electric charge with respect to the Kaluza-Klein
gauge field. Once the $y$-direction is taken to be compact,
$y \simeq y + 2 \pi R$, the parameter
$Q$ is discrete, $Q = \ft{\wh{Q}}{R}$.

We have to take $K$ to be independent of $y$ in order to compactify
this direction. If $D=4$, the transverse harmonic function $K$
takes the form $K \sim \log r$ and diverges for $r \rightarrow \infty$.
The solution is not asymptotically flat. This is typical for 
brane-like solutions, where the number of transverse dimensions
is smaller than three. Examples are seven-branes in ten dimensions,
cosmic strings in four dimensions and black holes in three dimensions.

\end{Solution}

\end{appendix}


\end{document}